\pdfoutput=1
\documentclass[11pt]{article}

\usepackage[letterpaper,margin=1in]{geometry}

\usepackage[utf8]{inputenc}
\usepackage[T1]{fontenc}
\usepackage{lmodern}
\usepackage{microtype}

\usepackage{amsmath,amssymb,amsfonts}

\usepackage{graphicx}
\usepackage{xcolor}
\usepackage{booktabs}
\usepackage{multirow}
\usepackage[section]{placeins}
\usepackage{array}
\usepackage{tabularx}

\usepackage{tikz}
\usetikzlibrary{arrows.meta,positioning,calc,shapes.geometric,backgrounds,fit}
\usepackage{pgfplots}
\pgfplotsset{compat=1.18}

\usepackage{listings}
\usepackage{enumitem}
\usepackage{etoolbox}
\usepackage{authblk}
\usepackage{xurl}
\usepackage[hidelinks]{hyperref}
\usepackage{cleveref}

\newcommand{\geos}{\textsc{GEOS}}
\newcommand{\langgraph}{\textsc{LangGraph}}
\newcommand{\code}[1]{\texttt{\small #1}}
\newcommand{\verdictpass}{\textsc{pass}}
\newcommand{\verdictwarn}{\textsc{warn}}
\newcommand{\verdictfail}{\textsc{fail}}
\newcommand{\verdictskip}{\textsc{skip}}
\newcommand{\verdictincomplete}{\textsc{incomplete}}

\title{\bfseries Reliability by Architecture:\\
Agentic Workflows for Validated Multi-Physics Simulation from Natural Language}

\author[1]{S\o{}ren Taverniers\thanks{Corresponding author:
\texttt{sorentav@stanford.edu}.}}
\author[1,2]{Eric Darve}
\author[3]{Hamdi A. Tchelepi}
\affil[1]{Department of Mechanical Engineering, Stanford University, Stanford, CA 94305, USA}
\affil[2]{Institute for Computational and Mathematical Engineering, Stanford University, Stanford, CA 94305, USA}
\affil[3]{Department of Energy Science \& Engineering, Stanford University, Stanford, CA 94305, USA}

\date{}

\begin{document}
\maketitle

\begin{abstract}
\noindent
{Preparing, executing, and validating strongly coupled multi-physics simulations
require expert time. Agentic workflows use large language models (LLMs) to coordinate
them from natural-language descriptions, but reliability requires checks of physical
consistency and fidelity to the problem description. Our agentic harness implements explicit task allocation:
LLMs interpret user requests and simulator diagnostics, generate input components
not covered by predefined construction rules, and propose repairs, while deterministic routines construct
supported components and check inputs, edits, and results. We evaluate CO$_2$-storage
problems using the multi-physics simulator GEOS, from compositional flow to flow--poromechanics with frictional fault
contact. In 80 attempts on two flow problems, grounding raises the proportion of structurally valid inputs from $5/20$
to $20/20$, but acceptance reaches $20/20$ only with complete rule-based input preparation.
With identical inputs, local repairs correct
three seeded defects while preserving valid inputs. Supporting tests identify
simulations passing physics checks despite differing from the problem description or reference
results.}
\end{abstract}

\section{Introduction}
\label{sec:motivation}

\subsection{Problem setting}

{In geologic CO$_2$ storage, multi-physics simulations of coupled subsurface flow and
geomechanics inform regulatory assessments~\cite{epa2010classvi} and analyses of
injection-induced deformation, as at In Salah~\cite{ringrose2013insalah}.}

{A multi-physics simulation requires setting up physical models,
geometry, material properties, initial and
boundary conditions where applicable, and numerical settings. These choices may be expressed,
for example, in \emph{configuration files}, as in the open-source simulators GEOS for subsurface
engineering applications~\cite{geosinputfiles,settgast2024geos} and OpenFOAM for computational
fluid dynamics (CFD)~\cite{openfoamcasefiles}. Material and numerical parameters must be consistent,
the mesh must be compatible with the selected discretization and appropriately represent
any physical interfaces present, and constitutive models, initial and boundary conditions,
and coupled solver settings must be mutually consistent.}

{Once the setup is assembled, running the simulation can require repeated
attempts, with manual diagnosis and correction of failures. Diagnostics from parsing,
initialization, or solution may not identify the modeling choice responsible. Even if the
simulation executes successfully, the computed solution may
not satisfy all relevant physical constraints; a completed calculation may also represent a different
physical model, initial state, or operating control from the one requested.}

{Checking the setup and simulated response for physical consistency and
agreement with the intended experiment therefore requires expertise in both the physical
models and the simulator's input and solver requirements. These labor-intensive workflows place high demands
on valuable expert time.}
{Agentic frameworks, which combine large language models (LLMs) with external
tool use~\cite{yao2023react,wu2024autogen}, offer a way to alleviate this burden by assisting
with simulation setup, execution, and validation.}

\subsection{Prior approaches and remaining challenges}
\label{sec:relatedwork}

LLM agents combine language-model reasoning with retrieval, external tools, and stateful
orchestration~\cite{yao2023react,schick2023toolformer,wang2023voyager,wu2024autogen,langgraph}.
{An agent uses an LLM to interpret a task and select actions or tool calls.}
{LLM-assisted simulation workflows} {have applied} these capabilities to CFD, porous-media flow, reactor thermal
hydraulics, atomistic simulation, and mechanics and geomechanics
\cite{chen2024metaopenfoam,lie2026agentic,autosam2026,genius2026,mcpsim2026,heo2026,bekele2025geosimai}.
{Across these studies, documentation and examples guide input generation, static checks
  assess generated code, and simulator execution provides diagnostics for repair and validation.
  Outside simulation-specific workflows, LLM refinement methods use self-critiques or reflections on previous attempts to
  guide subsequent responses~\cite{madaan2023selfrefine,shinn2023reflexion}, while software-agent
  methods use executable tests to assess repairs~\cite{jimenez2024swebench,yang2024sweagent}.}

{We identify four recurring failure modes that arise when LLMs
  generate or revise simulation inputs.}
{First, the LLM can generate input components that the simulator cannot accept. Retrieval of
documentation, examples, or templates can reduce obsolete syntax and missing-file errors
\cite{lewis2020rag,chen2024metaopenfoam,autosam2026}. However, supplying this context does not
enforce the input requirements: an LLM-written component may still {use unsupported syntax},
omit required structure, or refer to an undefined object.}

{Second, the LLM can combine individually valid components into a physically inconsistent setup
or one that differs from the request. Initial conditions, constitutive models, and operating
controls may be incompatible; an admissible value or material may be assigned to the wrong
region, or a schedule may differ from the requested operation. These errors concern the
relationships between components and their intended use. {Documentation and structural checks
can establish which input constructs are permitted without verifying all relationships between
them or agreement with the request.}}

{Third, an LLM-proposed correction can disrupt parts of the input that were already working.
  Execution-grounded simulation workflows use {simulator diagnostics} to revise generated inputs
  \cite{genius2026,lie2026agentic}. If those corrections are not constrained, however, they may modify
  valid parts of the input, remove a requested feature to avoid the failure, or repeat an edit
  that has not resolved the error.}

{Fourth, errors in a setup generated or repaired by the LLM can remain undetected even when
the simulation completes. \textsc{JutulGPT}~\cite{lie2026agentic} uses simulator-enforced {physics}
checks and verifies selected user requirements in porous-media flow. Strongly-coupled problems
combining flow and geomechanics place greater demands on validation: checks must address
consistency across interacting processes and constitutive models, and fidelity to more complex
problem descriptions.}

{These failure modes motivate an explicit allocation of tasks between LLM reasoning and
  deterministic code. Hybrid methods delegate numerical subproblems to external
  solvers~\cite{heo2026}, while {\textsc{SimulCost} reports that LLM-based parameter tuning can be
  slower than conventional scans in its main comparison across 11 simulators, with an additional
  production code assessed separately}~\cite{simulcost2026}. This motivates
  distinguishing operations that benefit from LLM reasoning from construction, calculation,
  and checking procedures that can be implemented in deterministic code.}

\subsection{\texorpdfstring{{Our contribution}}{Our contribution}}
\label{sec:approach}

{We develop an agentic harness that coordinates LLM interpretation and generation,
deterministic input construction, simulator execution, checks, and controlled repair.}
{Deterministic routines construct input
components covered by implemented construction rules and check the assembled inputs and
simulation results. The LLM interprets
problem descriptions and simulator diagnostics, generates input components not constructed
by deterministic routines, and proposes edits when no predefined correction applies.
{The agentic harness} constrains LLM-proposed repairs to limit changes to working
inputs and requires successful execution, applicable physics checks, and agreement with the
requested experiment before accepting a simulation.}

{We use \geos{}, an open-source simulator for coupled subsurface flow and
geomechanics~\cite{settgast2024geos}, as the simulation engine. GEOS has been used in real-world
case studies, such as at the Newberry enhanced geothermal system demonstration site, where it
coupled thermal, hydraulic, and mechanical processes to predict microseismic activity for
comparison with field
observations~\cite{sherman2017newberry}. Other field-based applications include modeling
CO$_2$-injection-induced fault opening at In Salah~\cite{dufour2025insalah} and hydraulic-fracture
growth at the Hydraulic Fracturing Test Site~1 (HFTS-1) in the Permian
Basin~\cite{glubokovskikh2023hfts}. GEOS's combination of fluid transport,
rock deformation, and fault mechanics capabilities makes this simulator a suitable testbed for our
{agentic harness}, which we demonstrate on CO$_2$-storage examples
ranging from compositional flow, which tracks fluid components across phases, to coupled
flow--poromechanics with frictional fault contact.}

{We assess input construction, coupled simulation capability, and repair through
distinct evaluations. An 80-attempt comparison examines how grounding and
deterministic preparation affect the fraction of accepted simulations. A separate contact
demonstration examines the construction, execution, and validation of a coupled setup.
Repair is evaluated through paired runs starting from identical copies of an input, with
repair disabled in one run and enabled in the other. These comparisons assess whether local
repairs correct specified configuration defects while preserving valid content. Supporting
tests examine errors that remain undetected by successful execution or conservation checks
alone. Together, these evaluations assess how task allocation and explicit acceptance
requirements affect the reliability of LLM-assisted simulation.}

\section{Methodology}
\label{sec:methodology}

\subsection{Workflow implementation}
\label{sec:hybrid}
\label{sec:arch}

{The agentic harness orchestrates simulation setup,
execution, validation, and repair around GEOS. {Deterministic routines in the agentic harness} prepare and
correct inputs, invoke GEOS, and check inputs and outputs; GEOS performs the numerical
solution of the governing equations. {We implement the agentic harness} by converting
a manual GEOS workflow---preparing the mesh and input deck, running the simulator, and examining
the results---into an LLM-assisted workflow orchestrated by \langgraph{}~\cite{langgraph}.}
{GEOS represents the simulation setup
through XML input files, referred to here as \emph{decks}. Their elements and attributes
specify the physical and numerical configuration. The GEOS XML schema specifies the allowed
input elements and attributes, where they may appear, and their required formats.}

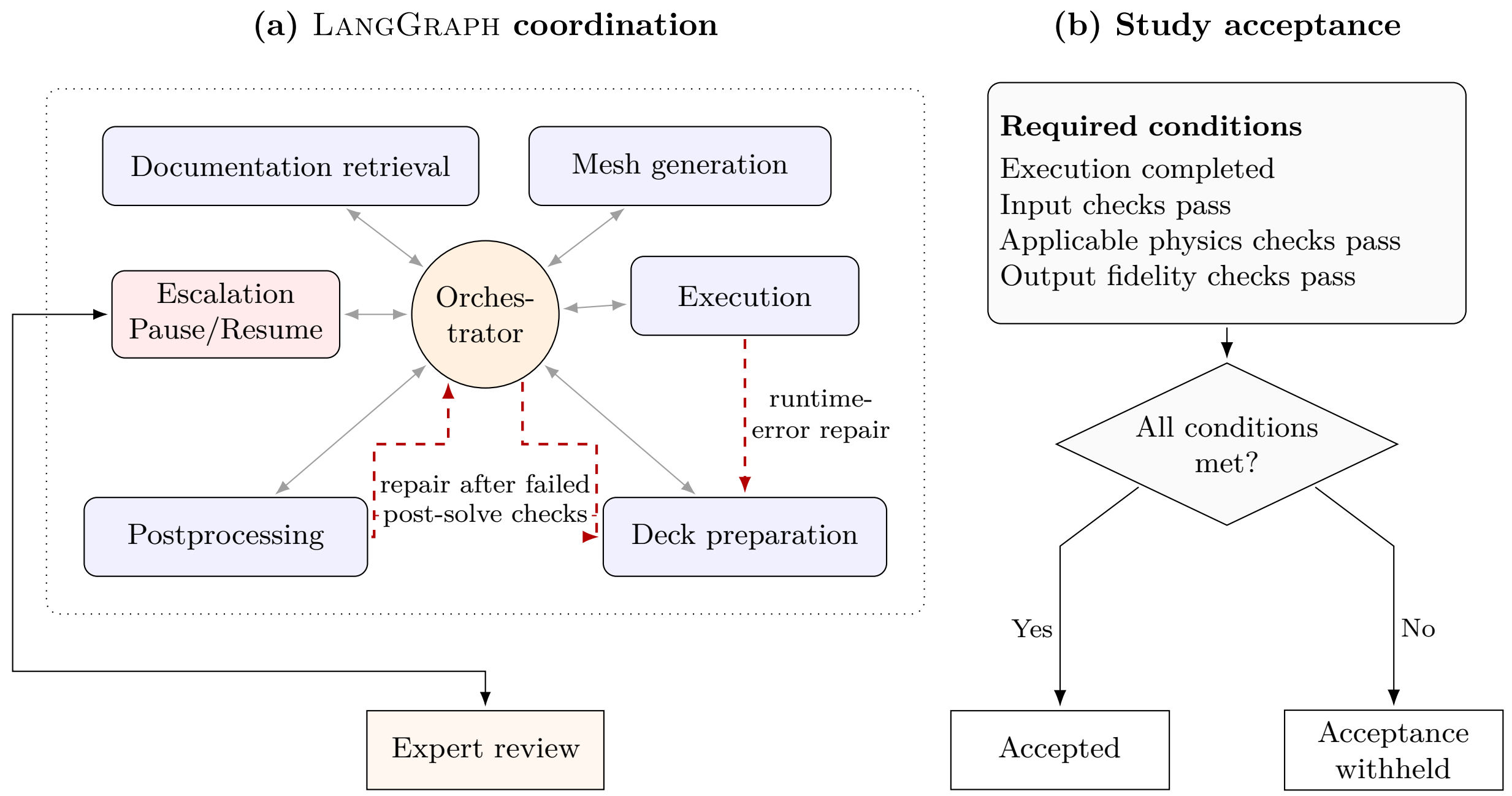
\begin{figure}[t]
  \centering
  \begin{tikzpicture}[
      x=1mm,y=1mm,font=\footnotesize,
      every node/.style={text=black},
      worker/.style={draw=black,rounded corners,text width=22mm,minimum height=8.5mm,fill=blue!6,align=center},
      hub/.style={draw=black,circle,minimum size=15mm,fill=orange!12,align=center},
      review/.style={worker,fill=red!8},
      expert/.style={draw=black,text width=23mm,minimum height=8.5mm,fill=orange!6,align=center},
      outcome/.style={draw=black,text width=21mm,minimum height=8.5mm,align=center},
      edge/.style={{Latex[length=1.6mm]}-{Latex[length=1.6mm]},gray!75,shorten <=1pt,shorten >=1pt},
      arrow/.style={-{Latex[length=1.8mm]},black,shorten <=1pt,shorten >=1pt},
      loop/.style={arrow,red!70!black,thick,dashed},
      elbl/.style={font=\scriptsize,fill=white,inner sep=1pt,align=center},
    ]
    \node[font=\small] at (0,31) {\textbf{(a) \textnormal{{\langgraph{}}} coordination}};
    \node[hub] (orch) at (0,0) {Orches-\\trator};
    \node[worker,text width=38mm,text=black] (query) at (-21,16) {\mbox{Documentation retrieval}};
    \node[worker,text width=30mm,text=black] (mesh) at (21,16) {\mbox{Mesh generation}};
    \node[worker,text width=28mm,text=black] (xml) at (28,-24) {\mbox{Deck preparation}};
    \node[worker] (run) at (28,2) {Execution};
    \node[worker,text width=28mm,text=black] (post) at (-28,-24) {Postprocessing};
    \node[review,text=black] (hr) at (-28,0) {Escalation\\\mbox{Pause/Resume}};
    \node[expert,text=black] (user) at (0,-47) {Expert review};

    \foreach \n in {query,mesh,xml,run,post,hr}{
      \draw[edge] (orch) -- (\n);
    }

    \draw[loop] (run.south) -- (run.south |- xml.north);
    \node[elbl,anchor=west,inner sep=0.5pt] at (28.5,-11) {runtime-\\error repair};
    \coordinate (physicalIn) at (-12,-24);
    \coordinate (physicalOut) at (12,-24);
    \coordinate (physicalInShoulder) at (-12,-14);
    \coordinate (physicalOutShoulder) at (12,-14);
    \draw[loop] (post.east) -- (physicalIn) -- (physicalInShoulder)
      -| (orch.240);
    \draw[loop] (orch.300) |- (physicalOutShoulder)
      -- (physicalOut) -- (xml.west);
    \node[elbl,inner sep=1pt,text=black] at (0,-18.5) {repair after failed};
    \node[elbl,inner sep=1pt,text=black] (physicalRepair) at (0,-21.7) {post-solve checks};
    \draw[red!70!black,thin] (physicalRepair.west) --
      (physicalIn |- physicalRepair.west);
    \draw[red!70!black,thin] (physicalRepair.east) --
      (physicalOut |- physicalRepair.east);
    \draw[edge,black] (hr.west) -- (-51,0) -- (-51,-38.5) -| (user.north);

    \begin{scope}[on background layer]
      \coordinate (workflowRight) at (42.5,0);
      \node[draw=black,dotted,rounded corners,
        fit=(query)(mesh)(xml)(run)(post)(hr)(orch)(workflowRight),inner sep=4mm] {};
    \end{scope}

    \node[font=\small] at (80,31) {\textbf{(b) Study acceptance}};
    \node[draw=black,rounded corners,text width=49mm,minimum height=26mm,
      align=left,fill=gray!4] (conditions) at (80,12)
      {\textbf{Required conditions}\\[2pt]
       Execution completed\\
       Input checks pass\\
       Applicable physics checks pass\\
       {Output fidelity checks pass}};
    \node[draw=black,diamond,aspect=2.1,align=center,inner sep=2pt,fill=gray!4]
      (decision) at (80,-14) {All conditions\\met?};
    \node[outcome] (accepted) at (62,-47) {Accepted};
    \node[outcome] (blocked) at (98,-47) {Acceptance\\withheld};
    \draw[arrow] (conditions.south) -- (decision.north);
    \draw[arrow] (decision.south west) -- (62,-25) --
      node[midway,left=1pt,elbl]{Yes} (accepted.north);
    \draw[arrow] (decision.south east) -- (98,-25) --
      node[midway,right=1pt,elbl]{No} (blocked.north);
  \end{tikzpicture}
  \caption{{Agentic harness coordination and study acceptance.
    (a) The dotted boundary encloses \langgraph{} nodes.
    Gray double-headed arrows summarize coordination and
    shared-state exchange. Deck preparation includes pre-solve admissibility and
    input-fidelity checks. Postprocessing includes physics checks and applicable input-
    and output-fidelity checks. Fidelity checks compare inputs and outputs with the user request.
    The escalation checkpoint manages pausing and resumption for
    expert review {outside the agentic harness}. {Red dashed arrows show optional repair
    after failed postprocessing checks through the orchestrator, including physics and fidelity
    failures, and runtime-error repair from execution to deck preparation.}
    (b) Study acceptance requires satisfying all listed conditions, including any additional
    fidelity checks performed outside the \langgraph{} workflow as part of the study.
    Panel~(b) shows assessment logic and outcomes, not additional
    \langgraph{} nodes (\Cref{sec:admissibility,sec:physics,sec:fidelity}).}
  }
  \label{fig:graph}
\end{figure}

{\langgraph{} executes documentation retrieval, mesh generation, deck preparation,
execution, and postprocessing as nodes in a workflow graph (\Cref{fig:graph}). Deck preparation
includes initial generation when no deck is supplied. Each node may
combine LLM calls, deterministic routines, and external tool use. Nodes read and update a
shared state containing the problem description, input decks, stage status, and records of
execution and validation. The orchestrator node serves as a coordination hub, with attached
deterministic routing rules selecting subsequent stages from workflow status, errors, check
results, and retry counts. \langgraph{} executes these transitions and feedback loops. Within the
orchestrator node, a locally hosted Llama-3 8B model generates commentary on workflow progress
that is retained in the shared state.}

{Input checks, including input-fidelity checks, operate during deck
preparation. The postprocessing node applies physics checks and applicable input- and
output-fidelity checks. Additional fidelity checks are performed by separate evaluation scripts outside the
\langgraph{} workflow.
The escalation checkpoint pauses the workflow for
unresolved input or runtime issues. Expert review takes place {outside the agentic harness};
the checkpoint enables resumption after intervention.
\Cref{sec:validation} describes the checks and final acceptance. \Cref{tab:split} summarizes
the allocation of tasks between LLM reasoning and {deterministic routines in the agentic harness}.}

{This sequence implements an interpret--act--validate loop, as used by
  \textsc{JutulGPT}~\cite{lie2026agentic} and related reasoning-and-action
  methods~\cite{yao2023react}. {\Cref{sec:app:nodes} describes the shared state and node interfaces.}}

\begin{table}[!htbp]
  \centering

  \caption{{Allocation of responsibilities between LLMs and deterministic routines
    in the agentic harness surrounding GEOS.} The simulator numerically solves the governing equations.}
  \label{tab:split}
  \small\renewcommand{\arraystretch}{1.15}
  \begin{tabularx}{\textwidth}{@{}>{\raggedright\arraybackslash}p{0.17\textwidth}>{\raggedright\arraybackslash}X>{\raggedright\arraybackslash}X@{}}
    \toprule
    Stage & LLM contribution & {Deterministic routines} \\
    \midrule
    Retrieve and interpret & {Use repository browsing and search tools to find reference decks when embedding retrieval is unavailable or returns weak matches} & {Retrieve decks by embedding similarity; rank candidates against the requested physics; assemble current-schema context} \\
    Mesh and deck construction & Choose geometry strategy and tools; {write components not synthesized by deterministic code} & Size the mesh; map regions and materials; {convert units; correct links between deck objects; construct input components from requested parameters using predefined rules} \\
    {Physical setup} & Extract requested parameters & Calculate initial states and controls; {make referenced input files available to the simulator}; check admissibility and input fidelity \\
    Runtime repair & {Diagnose the failure from the supplied log excerpts and propose a local edit when no prescribed correction applies} & {Select error messages and surrounding log lines for LLM diagnosis}; apply specified corrections; check schema compliance and preservation of required, previously valid deck content; enforce retry limits \\
    {Execution and postprocessing} & {Optionally generate a narrative summary of the simulation results and the outcomes of the post-solve checks} & {Invoke GEOS; extract execution information and numerical quantities to check completion and apply physics, input-fidelity, and output-fidelity checks} \\
    \bottomrule
  \end{tabularx}

\end{table}

\subsection{Constructing and checking the physical setup}
\label{sec:boundary}

{{The} generated-deck path starts from an LLM-generated {GEOS input deck}; the supplied-deck path starts
  from a copy of an existing input. Before execution, deterministic routines prepare the deck
  for the mesh and requested physical setup, then check it. We call this stage
  \emph{deterministic preparation}. Its input checks also apply after repair changes a deck.}

\subsubsection{LLM interpretation and grounded deck generation}
\label{sec:grounding}

{The GEOS input schema evolves between releases. In the schema snapshot used
in this study, our parser counts roughly $275$ element types and nearly $3{,}000$ attributes.
Such an evolving schema can lead to plausible-looking input decks that use obsolete syntax.}

In the generated-deck path, the LLM extracts requested geometry, phases, controls, and simulation
time. Deterministic code builds components for which explicit constructors are available, and the
LLM writes the remaining XML. The supplied-deck path skips initial generation and operates on a
copy of the supplied input.

{Grounding supplies the deck-generation LLM with current GEOS element documentation,
  retrieved reference decks~\cite{lewis2020rag}, and fixed construction instructions.
  The documentation explains valid attributes and their meanings; reference decks show working
  combinations. The construction instructions encode additional constraints identified during
  development of the framework (\Cref{sec:app:det}). This context guides generation but does
  not enforce input requirements.}

\subsubsection{Deterministic normalization and synthesis}
\label{sec:construction}

{Deterministic preparation combines normalization and synthesis.
Normalization corrects missing or inconsistent declarations, object references, and
mesh-region assignments in a candidate deck, and converts units where required.
Synthesis calculates dependent quantities and constructs input components from requested
parameters, such as well completions and timed injection sources. For example, it calculates
hydrostatic pressure using the selected fluid model's density and sets bottom-hole pressure
(BHP) for pressure-controlled wells relative to formation pressure at the specified downhole
reference elevation. This keeps initial states and well controls consistent with the fluid
model without LLM estimates.
\Cref{sec:app:det,sec:app:contact} give the deterministic preparation rules and the contact-specific
construction and initialization procedure.}

{When a request leaves a choice unspecified, deterministic preparation applies a {predefined
default} or stops for clarification if it recognizes the ambiguity and no rule applies.
Synthesis records each default it supplies, including the component, field, value, and reason.
This assumption log does not include LLM-generated values, whose origin is not identified in
the returned XML. Unspecified choices may still go unnoticed if neither LLM interpretation
nor a preparation rule identifies them (\Cref{sec:limitations}).}

{Encoding preparation in deterministic routines supports reproducibility.
{With the same rules and software versions, identical candidate input decks and
parameters undergo the same deterministic transformations.} Subsequently, the saved final deck, mesh,
companion files, and execution settings can be used to rerun the simulation with an identical
setup. The calculations and input modifications that produced this setup can be inspected in
the preparation scripts.}

\subsubsection{Pre-solve admissibility and input fidelity checks}
\label{sec:admissibility}

{XML schema validation checks whether the assembled deck uses the required elements and
  allowed attributes. Passing this structural check does not establish physical admissibility
  or agreement with the request.}

Pre-solve admissibility checks ask whether the assembled input satisfies simulator and physical
constraints. They test both individual bounds, such as positive solid density, and relationships
between components, such as compatible saturation parameters and solver-region assignments.

Input fidelity checks ask a different question: does that admissible setup match the request?
A plausible permeability assigned to the wrong region, for example, requires a comparison with
the intended material assignment rather than a scalar admissibility bound.

{Each check runs when its required inputs are available. A detected mismatch prevents
  acceptance. If an implemented deterministic correction specifies the required change, it is
  applied to the candidate deck; otherwise, the input and {check result} are retained for human
  review. The workflow does not silently replace a requested value with an admissible alternative.}

{The input-check rows of \Cref{tab:validation-roles} summarize these distinctions.
  {\Cref{sec:app:bands} summarizes the implemented input checks.}}

\subsection{Controlled local repair}
\label{sec:selfheal}
\label{sec:guards}

{During framework development, regenerating a whole deck to remove event warnings also removed
the event that drove the solver. The simulator exited without taking a single time step. Correcting
the reported error had removed the requested calculation. The repair paths therefore use the
existing deck as the basis for correction and check the candidate's required input structure and
physical setup. The supplied-deck path restricts each proposal to a local edit.}

{After a diagnosed failure, the workflow first considers an applicable, fully-specified
  deterministic correction. Otherwise, the LLM receives the request, current deck, and {simulator diagnostics}
  and proposes a local edit. {In the supplied-deck path, deterministic code applies
  the proposal to a copy of the supplied input deck and checks the schema, input constraints,
  and preservation of required working content before another solve.}
  A rejected candidate leaves the current deck unchanged.}

{We call this \emph{controlled local repair} because both the permitted edit scope and
  the number of correction attempts are limited. The workflow does not repeat an edit when the
  same failure recurs at the same element or attribute. It may try another permitted edit until
  reaching the retry limit, after which an unresolved input or runtime failure is sent for human
  review. \Cref{sec:app:repair} gives the permitted edit forms and retry limits.}

{Post-solve repair uses feedback from the postprocessing node. {When automatic
  repair is enabled, a \verdictfail{} report with diagnostic feedback can route the workflow
  back to deck preparation to address the failed physics or fidelity checks.} This loop permits at most two retries.
  A \verdictwarn{} or \verdictincomplete{} verdict withholds acceptance without initiating
  repair (\Cref{sec:physics,sec:app:bands}).}
{Corrections identified during framework development are incorporated into
  deterministic preparation in later framework configurations when they can be expressed as
  predefined rules.}

\subsection{Post-solve validation and final acceptance}
\label{sec:validation}

{The problem description specifies the intended physical scenario, physical laws,
parameters, and requirements.
{The agentic harness} constructs the simulation, applies physics checks, and uses input and output
fidelity checks to assess agreement with that description. A \emph{physics gate} assesses the
computed physical behavior; {input- and output-fidelity checks compare the assembled inputs
and saved results, respectively, with the request or a specified reference.} These roles are summarized in
\Cref{tab:validation-roles}; study acceptance is distinct from graph coordination, as shown in
\Cref{fig:graph}.}

\begin{table}[!htbp]
  \centering

  \caption{{Questions answered by the validation stages. Input- and output-fidelity
  checks assess agreement with the request or a specified reference, rather than {physical
  consistency}.}}
  \label{tab:validation-roles}
  \small\renewcommand{\arraystretch}{1.15}
  \begin{tabularx}{\textwidth}{@{}>{\raggedright\arraybackslash}p{0.21\textwidth}>{\raggedright\arraybackslash}X>{\raggedright\arraybackslash}X@{}}
    \toprule
    Check & Question & Example \\
    \midrule
    Schema validation & Does the deck use the accepted input structure? & Required elements and allowed attributes \\
    Pre-solve admissibility & Are values and component relationships physically and numerically admissible? & Positive solid density; compatible solver-region assignments \\
    Input fidelity & Does the assembled setup agree with the request? & Requested injection rate and stopping time \\
    Post-solve physics gate & Does the computed behavior satisfy the applicable {physics} checks? & {Increase in stored CO$_2$ matches the injected mass} \\
    Output fidelity & {Do saved results satisfy the requested output requirements and agree with any specified reference?} & CO$_2$ inventory after injection stops and requested simulation duration; baseline-trajectory agreement \\
    \bottomrule
  \end{tabularx}

\end{table}

\subsubsection{Post-solve physics gate}
\label{sec:physics}

{The physics gate checks quantities extracted from the mesh and saved fields for
  conservation, admissible ranges, expected trends, and consistency between coupled processes.
  The simulated physics determines which implemented checks apply; for example, mechanics checks
  are not needed for a flow-only problem.}

{The check library and the rules for selecting and reporting checks are reused across the
  flow, poromechanics, contact, and reference-deck cases. {\Cref{tab:physics,sec:app:bands} summarize the check categories and applicability policy.}}

{The gate reports \verdictpass{} only if every applicable check passes. A check that does not
  apply is recorded as \verdictskip{}. Missing data, an error while evaluating a required check,
  a warning, or a failed check prevents acceptance. \Cref{sec:app:bands} gives the verdict rules.}

\subsubsection{Output fidelity checks and final acceptance}
\label{sec:fidelity}

{Passing physics checks does not establish that a simulation matches the user request.
For example, in a closed domain, a mass-conservation check compares the increase in stored
CO$_2$ with the amount injected in the simulation. This check can pass even if the injection
rate differs from the user request. Input-fidelity checks compare the encoded rate and time window with the
request; output-fidelity checks compare the resulting CO$_2$ mass, simulation duration,
and saved-output coverage with it.}

{The postprocessing node applies physics checks and applicable input- and output-fidelity checks.
Failures can trigger controlled local repair and rerunning when automatic repair is enabled and
retries remain. Additional fidelity checks performed by separate evaluation scripts
outside the \langgraph{} workflow do not trigger workflow repair.}

{Final acceptance of a simulation result requires completed execution, a physics-gate \verdictpass{},
and passing all required input- and output-fidelity checks. Reliability here means withholding
acceptance when a required check fails or evidence is missing. Saved inputs, simulator outputs,
and check reports make this decision traceable. The injection-schedule and trajectory tests
illustrate these criteria (\Cref{sec:res:guards}).}
{Passing the specified checks does not establish that the tested scenarios cover
the practical problem of interest; experts must assess that applicability before drawing
scientific conclusions.}

\section{Evaluation design}
\label{sec:evaluation}

{We simulate four physical scenarios with GEOS: an emplaced CO$_2$ plume (T1),
a two-well flow case (T2), and two flow--poromechanics cases. The continuum fault-zone case (T3)
represents faults as finite-width regions within the porous rock and couples fluid pressure
to rock deformation. The frictional-contact case (T4) instead represents faults as discrete
surfaces, with a contact model describing relative motion across them. \Cref{fig:domains} shows the simulation domains, and
\Cref{tab:cases} provides the scenario specifications. \Cref{tab:incremental} identifies which scenarios
test setup, coupled capability, and repair.}

\definecolor{clSeal}{RGB}{176,196,222}
\definecolor{clSand}{RGB}{233,205,150}
\definecolor{clFault}{RGB}{120,120,124}
\definecolor{clCO2}{RGB}{0,114,178}
\definecolor{clWell}{RGB}{213,94,0}
\tikzset{
  layerlbl/.style={font=\scriptsize\itshape, text=black!70},
  dimlbl/.style={font=\scriptsize, text=black!60},
  annot/.style={font=\scriptsize},
  wellln/.style={line width=1.1pt, clWell},
}

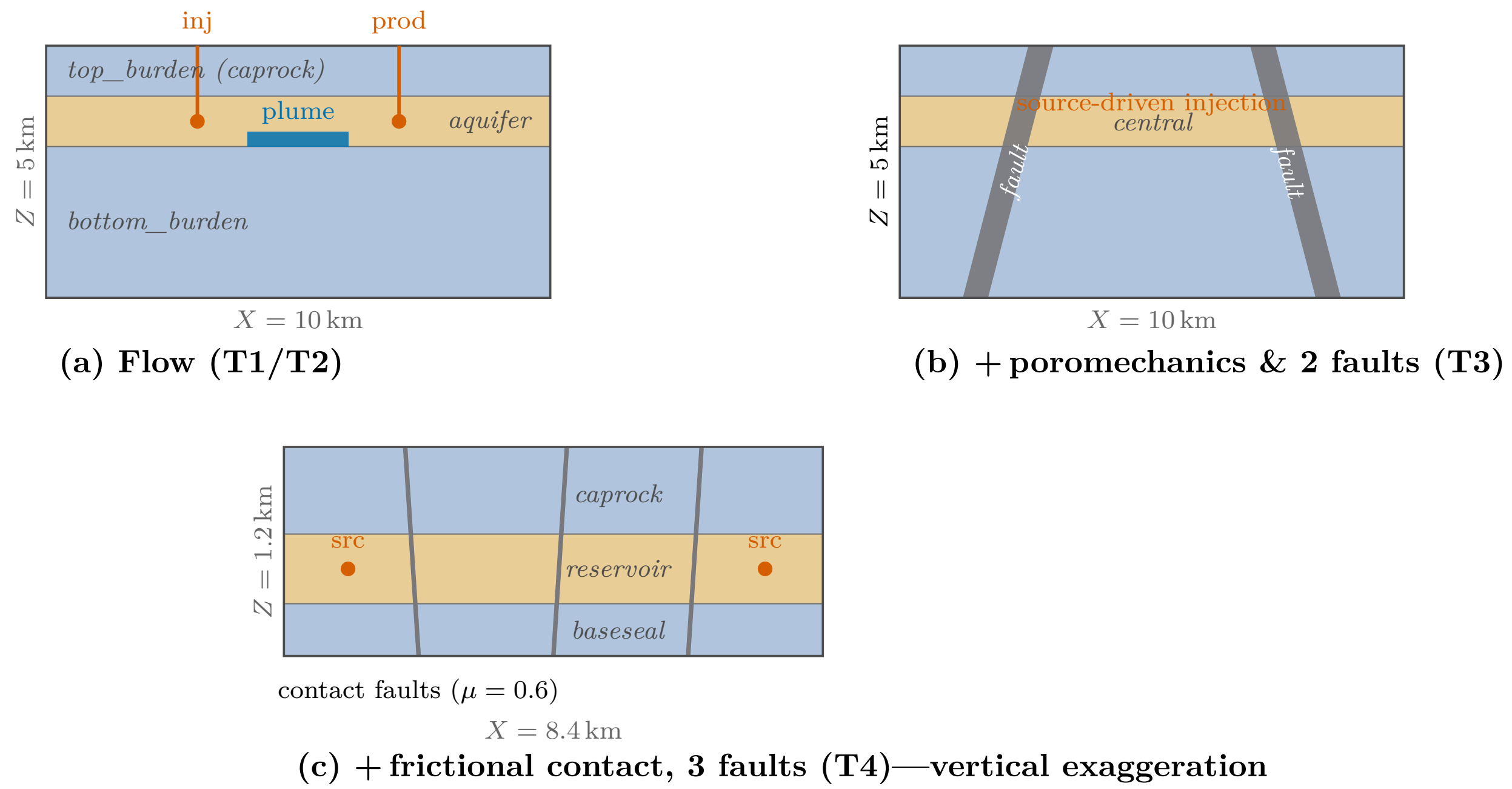
\begin{figure}[!t]
  \centering
  \begin{tikzpicture}[x=0.55cm, y=0.55cm]
    \def\W{10} \def\H{5}
    \fill[clSeal] (0,0) rectangle (\W,3);
    \fill[clSand] (0,3) rectangle (\W,4);
    \fill[clSeal] (0,4) rectangle (\W,5);
    \draw[black!55, line width=0.4pt] (0,3) -- (\W,3) (0,4) -- (\W,4);
    \draw[black!70, line width=0.7pt] (0,0) rectangle (\W,\H);
    \fill[clCO2, opacity=0.85] (4,3) rectangle (6,3.3);
    \node[annot, text=clCO2] at (5,3.68) {plume};
    \draw[wellln] (3,\H) -- (3,3.4);  \fill[clWell] (3,3.5) circle (0.8mm);
    \draw[wellln] (7,\H) -- (7,3.4);  \fill[clWell] (7,3.5) circle (0.8mm);
    \node[annot, text=clWell, anchor=south] at (3,\H) {inj};
    \node[annot, text=clWell, anchor=south] at (7,\H) {prod};
    \node[layerlbl, anchor=west] at (0.15,4.5) {top\_burden (caprock)};
    \node[layerlbl, anchor=east] at (\W-0.15,3.5) {aquifer};
    \node[layerlbl, anchor=west] at (0.15,1.5) {bottom\_burden};
    \node[dimlbl, below] at (\W/2,0) {$X=10$\,km};
    \node[dimlbl, rotate=90, anchor=south] at (0,\H/2) {$Z=5$\,km};
    \node[font=\small\bfseries, anchor=north west] at (0,-0.75) {{(a) Flow (T1/T2)}};
  \end{tikzpicture}
  \hfill
  \begin{tikzpicture}[x=0.55cm, y=0.55cm]
    \def\W{10} \def\H{5}
    \fill[clSeal] (0,0) rectangle (\W,3);
    \fill[clSand] (0,3) rectangle (\W,4);
    \fill[clSeal] (0,4) rectangle (\W,5);
    \draw[black!55, line width=0.4pt] (0,3) -- (\W,3) (0,4) -- (\W,4);
    \fill[clFault, opacity=0.9] (1.25,0) -- (1.75,0) -- (3.05,5) -- (2.55,5) -- cycle;
    \fill[clFault, opacity=0.9] (8.25,0) -- (8.75,0) -- (7.45,5) -- (6.95,5) -- cycle;
    \draw[black!70, line width=0.7pt] (0,0) rectangle (\W,\H);
    \node[annot, text=clWell] at (5,3.85) {source-driven injection};
    \node[layerlbl, text=white] at (2.3,2.5) {\rotatebox{75}{fault}};
    \node[layerlbl, text=white] at (7.7,2.5) {\rotatebox{-75}{fault}};
    \node[layerlbl] at (5,3.5) {central};
    \node[dimlbl, below] at (\W/2,0) {$X=10$\,km};
    \node[dimlbl, text=black, rotate=90, anchor=south] at (0,\H/2) {$Z=5$\,km};
    \node[font=\small\bfseries, anchor=north west] at (0,-0.75) {(b) +\,poromechanics \& 2 faults (T3)};
  \end{tikzpicture}

  \vspace{1.4em}
  \begin{tikzpicture}[x=0.70cm, y=1.9cm]
    \def\W{8.4}
    \fill[clSeal] (0,0) rectangle (\W,0.3);
    \fill[clSand] (0,0.3) rectangle (\W,0.7);
    \fill[clSeal] (0,0.7) rectangle (\W,1.2);
    \draw[black!55, line width=0.4pt] (0,0.3) -- (\W,0.3) (0,0.7) -- (\W,0.7);
    \draw[clFault, line width=1.4pt] (2.1,0) -- (1.89,1.2);
    \draw[clFault, line width=1.4pt] (4.2,0) -- (4.41,1.2);
    \draw[clFault, line width=1.4pt] (6.3,0) -- (6.51,1.2);
    \draw[black!70, line width=0.7pt] (0,0) rectangle (\W,1.2);
    \fill[clWell] (1.0,0.5) circle (0.8mm);
    \fill[clWell] (7.5,0.5) circle (0.8mm);
    \node[annot, text=clWell, anchor=south, font=\scriptsize] at (1.0,0.55) {src};
    \node[annot, text=clWell, anchor=south, font=\scriptsize] at (7.5,0.55) {src};
    \node[layerlbl] at (5.2,0.92) {caprock};
    \node[layerlbl] at (5.2,0.5) {reservoir};
    \node[layerlbl] at (5.2,0.15) {baseseal};
    \node[annot, anchor=north] at (2.1,-0.06) {\scriptsize contact faults ($\mu=0.6$)};
    \node[dimlbl, below] at (\W/2,-0.30) {$X=8.4$\,km};
    \node[dimlbl, rotate=90, anchor=south] at (0,0.6) {$Z=1.2$\,km};
    \node[font=\small\bfseries, anchor=north west] at (0,-0.48) {(c) +\,frictional contact, 3 faults (T4)---vertical exaggeration};
  \end{tikzpicture}

  \caption{{Simulation} domains of the four scenarios simulated in GEOS, as $X$--$Z$ cross-sections (not to
    a common scale; all cases place a storage aquifer between seals, \Cref{tab:cases}).
    {Three principal evaluation designs reuse these scenarios, as summarized in \Cref{tab:incremental}.}
    \textbf{(a)}~The
    flow cases share a layered aquifer between caprock and base seals. T1 starts with an
    emplaced mobile free-gas CO$_2$ plume whose upward migration is checked, whereas T2
    starts with 100\% brine and uses a bottom-hole-pressure-controlled injector/producer
    pair. The depicted initial plume belongs to T1 and the wells to T2. \textbf{(b)}~The
    coupled poromechanics case (T3) adds two steeply dipping ($\pm75^\circ$) continuum
    fault zones that compartmentalize the domain, with source-driven injection.
    \textbf{(c)}~{The SPE11C-inspired contact case (T4)~\cite{spe11}, on a
    {shallower section with a larger width-to-depth ratio}, treats three steep faults as discrete frictional surfaces,
    using Lagrange multipliers to enforce contact constraints and a friction coefficient
    of $\mu=0.6$. Injection uses two volumetric \code{SourceFlux} regions; the vertical axis
    is exaggerated for legibility.} Colors:
    sand~=~storage reservoir/aquifer, blue-gray~=~seal, gray~=~fault, orange~=~wells or
  sources, blue~=~emplaced CO$_2$.
  }
  \label{fig:domains}
\end{figure}

\suppressfloats[t]
\begin{table}[t]
  \centering

  \caption{The three principal evaluations. Physical scenarios are reused to answer different
  questions; generated and supplied inputs are kept distinct.}
  \label{tab:incremental}
  \small\renewcommand{\arraystretch}{1.15}
  \begin{tabularx}{\textwidth}{@{}>{\raggedright\arraybackslash}p{0.22\textwidth}>{\raggedright\arraybackslash}X>{\raggedright\arraybackslash}X@{}}
    \toprule
    Evaluation & Inputs and comparison & Question \\
    \midrule
    Generated-deck flow ablation & Plume (T1) and two-well flow (T2); four configurations, 10 attempts per case and configuration & Which setup components increase accepted yield? \\
    Contact capability demonstration & Natural-language description of flow--poromechanics with frictional contact (T4) & Can the workflow construct, execute, and validate this coupled setup? \\
    Paired supplied-deck repair & Two-well flow (T2) and continuum flow--poromechanics (T3); identical valid or defective decks with repair disabled and enabled & Can a local correction restore execution while preserving the required setup? \\
    \bottomrule
  \end{tabularx}

\end{table}

\subsection{Generated-deck flow ablation}

{The flow ablation successively adds grounding, normalization, and synthesis to ungrounded
  generation, giving four configurations. Each configuration has 10 plume attempts (T1) and 10
  two-well attempts (T2), for 80 attempts in total. Corresponding attempts request the same
  sampling seed across configurations (\Cref{sec:app:campaign}). For each configuration, we
  report how many generated decks satisfy the XML schema and how many attempts produce an
  accepted simulation {without human intervention. The fraction of attempts
  accepted in this way is termed \emph{accepted yield}}. For both T1 and T2, all four configurations use the same checks and
  acceptance criteria.}

{All configurations use Claude Sonnet~4.6 for deck and mesh generation. The ablation varies
  the grounding context supplied to the deck-generation LLM and the deterministic preparation
  applied to the generated deck, while keeping mesh generation unchanged. Grounding comprises
  documentation, reference decks, and construction instructions
  (\Cref{sec:grounding}). These are enabled or disabled as a package; their individual
  contributions are not isolated.}

{{When} deterministic synthesis is enabled, routines calculate the requested initial state
  and well controls using the selected fluid model.
  These include the hydrostatic brine background, plume free-gas saturation, and well pressures
  relative to formation pressure. When synthesis is disabled, the same checks assess the
  generated values without constructing replacements.}

{A comparison separate from the generated-deck ablation tests normalization using 20 grounded
  decks saved from this ablation. Each deck is run
  with synthesis enabled, once with normalization and once without it, using the same acceptance
  criteria and with runtime repair disabled (\Cref{sec:app:normalization-diagnostic}).}

\subsection{Contact capability demonstration}

{{The contact demonstration uses T4, a layered reservoir with three frictional faults
whose permeability depends on normal contact loading through a hydraulic-aperture relation.}
The fluid system is CO$_2$--brine at $333.15$\,K with a fixed
salinity of $2.5$\,mol NaCl per kilogram of brine. The natural-language description requests
30 days of injection through two source regions followed by 15 days of monitoring.}

{The problem description specifies geometry, material properties, physical laws,
initial loading, injection, and output requirements. {The agentic harness} supplies the GEOS-specific
model and solver configuration through its supported contact construction procedure,
generates a fresh mesh and deck, and executes and checks the coupled simulation.}
{The contact-case physics-gate and fidelity
checks cover the physical fields, contact behavior, and requested injection schedule.
\Cref{sec:app:contact} details the deterministic preparation and contact-specific checks.}

\subsection{Paired supplied-deck repair evaluation}

{The repair evaluation uses the two-well flow case (T2) and the continuum
  flow--poromechanics case (T3). T2 is a 365-day simulation with a
  bottom-hole-pressure-controlled injector and producer (\Cref{fig:domains}, panel~a). T3 injects
  $500$ million kg over $730$ days and then monitors
  for $182.5$ days in a mesh with two finite-width fault zones (\Cref{fig:domains}, panel~b).}

{For each setup, the paired evaluation compares two runs starting from identical copies
  of a supplied input deck and its companion files, with repair disabled in one run and enabled
  in the other. Using supplied inputs keeps generation errors out of the comparison.}

{We first test each valid baseline, then introduce the same specified defect into both copies
  and repeat the comparison. Both runs receive the same deterministic preparation and acceptance
  checks; only the repair-enabled run may receive a checked local correction after a failure.
  Together, these pairs test whether controlled local repair can correct known configuration defects
  without changing valid inputs.}

{In T2, the tests break a well-control reference or remove a relative-permeability table
whose values are stated in the request. In T3, the test breaks a solid-solver reference. These
three defects stop execution and are not corrected by the existing
deterministic preparation routines. They were selected to test known repair paths, rather than sampled to
estimate how often repair succeeds on arbitrary defects. \Cref{sec:app:repair-evaluation}
specifies the permitted edit scope and the checks used to ensure that other input-deck content
remains unchanged.}

\subsection{Supporting validation and portability tests}

{The supporting tests examine what can be missed by a completed solve or an individual
  check. \Cref{tab:supporting-designs} maps each validation question to its test;
  {\Cref{sec:res:guards} reports the outcomes, and \Cref{sec:appendix-evidence} gives the supporting methods and evidence.}}

\begin{table}[!htbp]
  \centering
  {
    \caption{Supporting validation tests and the questions they examine. The tests distinguish
      {physics} checks, fidelity to the request, and workflow responses to failures.}
    \label{tab:supporting-designs}
    \small\renewcommand{\arraystretch}{1.15}
    \begin{tabularx}{\textwidth}{@{}>{\raggedright\arraybackslash}X>{\raggedright\arraybackslash}p{0.26\textwidth}>{\raggedright\arraybackslash}X@{}}
      \toprule
      Question & Test & Evidence sought \\
      \midrule
      {How closely does a coupled solver configuration reproduce an analytical solution?} & {Single-phase Terzaghi consolidation, with separate mesh and timestep refinement} & {Pressure, displacement, and settlement errors from direct GEOS runs, without agents} \\
      Can a mass-balanced result violate the requested schedule? & Injection followed by monitoring & Input schedule and saved inventory compared with the request \\
      What can output checks miss? & Seven input defects, with pre-solve checks bypassed & Differences between output detection, input admissibility, and fidelity of material assignment \\
      Do the implemented check functions detect their intended violations? & Direct physics-gate check-library test & Detection and false alarms without simulation-file extraction \\
      {How closely do extracted inventories agree with published reference results?} & {SPE11B reference comparison} & {CO$_2$ inventory histories compared with the official \texttt{geos1} submission} \\
      {Does passing physics checks imply agreement with a reference trajectory?} & {Continuum numerical-parameter test} & {Physics verdicts compared with baseline-trajectory differences} \\
      Can a known incompatible request be escalated before execution? & Contact--well compatibility and graph-routing tests & Solver incompatibility distinguished from the workflow's routing response \\
      \bottomrule
    \end{tabularx}
  }
\end{table}

{Separate OpenFOAM tests assess portability of input preparation without using
  {the GEOS-based agentic harness}.}
{An optional pre-run LLM critic was evaluated separately and is excluded from the
  primary claims and accepted-yield comparison. {It is disabled in the principal evaluations
  and omitted from \Cref{fig:graph}.} \Cref{sec:app:critic} describes its evaluation
  and results.}

\section{Results}
\label{sec:results}

\subsection{Generated-deck flow ablation}
\label{sec:res:ablation}

{\Cref{fig:ablation} reports results for the plume (T1) and two-well (T2) benchmarks
  and separates the effects of grounding, synthesis, and normalization into three comparisons.
  Panel~(a) compares ungrounded and grounded generation with
  normalization and synthesis disabled. Grounding raises generation-stage schema validity
  from $5/20$ to $20/20$, but neither configuration produces an accepted simulation
  (\Cref{tab:ablation}).}

{Panel~(b) compares separately generated decks with and without
  deterministic synthesis, keeping grounding and normalization enabled. Without synthesis,
  the generated decks fail requirements for the requested initial state or well controls,
  leaving accepted yield at $0/20$. Deterministic synthesis calculates these quantities from
  the extracted parameters and selected fluid model, raising acceptance to $10/10$ for each
  benchmark and thus $20/20$ overall. {The two-sided 95\% Wilson interval is $83.9$--$100\%$.}
  \Cref{sec:app:campaign} reports which initial-state and well-control requirements were
  violated and how many attempts were affected in T1 and T2.}

{Panel~(c) shows a separate normalization comparison using
  identical saved grounded decks, with deterministic synthesis enabled in both runs.
  Normalization raises acceptance from $9/20$ to $20/20$ without turning any passing result
  into a non-passing one: T1 acceptance rises from $9/10$ to $10/10$, and T2 acceptance from
  $0/10$ to $10/10$. {Normalization supplies supporting structure required by the
  synthesized components: it removes a material-model component defined twice, adds missing
  regional pressure statistics, and corrects an invalid mesh-region layout.} \Cref{sec:app:normalization-diagnostic}
  details the failures and the normalization edits that resolve them.}

\begin{table}[!htbp]
  \centering
  {
    \caption{Flow-ablation outcomes, with 10 attempts per benchmark and configuration.
      Rows successively add components. Schema validity is assessed after generation and its allowed
      internal retries; acceptance additionally requires simulation completion, the physics gate,
    and {input- and output-fidelity checks}. {Intervals are two-sided 95\% Wilson intervals.}}
    \label{tab:ablation}
    \small\renewcommand{\arraystretch}{1.15}
    \begin{tabularx}{\textwidth}{@{}Xcccc@{}}
      \toprule
      Configuration & Schema-valid & \multicolumn{2}{c}{Accepted} & Combined accepted \\
      & (both cases) & Plume & Two wells & yield [95\% CI] \\
      \midrule
      Ungrounded generation & $5/20$ & $0/10$ & $0/10$ & $0/20$ [0--16.1\%] \\
      $+$ grounding & $20/20$ & $0/10$ & $0/10$ & $0/20$ [0--16.1\%] \\
      $+$ normalization & $20/20$ & $0/10$ & $0/10$ & $0/20$ [0--16.1\%] \\
      $+$ synthesis & $20/20$ & $10/10$ & $10/10$ & $20/20$ [83.9--100\%] \\
      \bottomrule
    \end{tabularx}
  }
\end{table}

\definecolor{abEnabled}{RGB}{0,114,178}
\begin{figure}[htbp]
  \centering
  \pgfplotsset{abContrast/.style={
      width=\linewidth, height=5.0cm,
      xmin=-0.6, xmax=1.6, ymin=0, ymax=130,
      xtick={0,1}, xticklabels={without, with},
      ytick={0,25,50,75,100},
      bar width=18pt,
      x tick label style={font=\small}, y tick label style={font=\footnotesize},
      ymajorgrids, grid style={gray!20}, axis line style={gray!60},
  }}
  \begin{minipage}[t]{0.32\textwidth}
    \centering
    {\small\bfseries (a) Grounding}\par
    {\small Schema-valid decks (\%)}\par
    \smallskip
    \begin{tikzpicture}
      \begin{axis}[abContrast]
        \addplot[ybar, bar shift=0pt, fill=gray!55, draw=none,
          error bars/.cd, y dir=both, y explicit,
          error bar style={black, semithick},
        error mark options={rotate=90, black, mark size=2.5pt}]
        coordinates {(0,25) += (0,21.9) -= (0,13.8)};
        \addplot[ybar, bar shift=0pt, fill=abEnabled, draw=none,
          error bars/.cd, y dir=both, y explicit,
          error bar style={black, semithick},
        error mark options={rotate=90, black, mark size=2.5pt}]
        coordinates {(1,100) += (0,0) -= (0,16.1)};
        \node[font=\small, anchor=south] at (axis cs:0,51) {$5/20$};
        \node[font=\small, anchor=south] at (axis cs:1,104) {$20/20$};
      \end{axis}
    \end{tikzpicture}\par
    {\footnotesize Separate generations\par
      \smallskip
      Plume: $5/10\rightarrow 10/10$\par
      Two wells: $0/10\rightarrow 10/10$\par
      \smallskip
    Normalization and synthesis\\disabled.\par}
  \end{minipage}\hfill
  \begin{minipage}[t]{0.32\textwidth}
    \centering
    {\small\bfseries (b) Synthesis}\par
    {\small Accepted yield (\%)}\par
    \smallskip
    \begin{tikzpicture}
      \begin{axis}[abContrast]
        \addplot[ybar, bar shift=0pt, fill=gray!55, draw=none,
          error bars/.cd, y dir=both, y explicit,
          error bar style={black, semithick},
        error mark options={rotate=90, black, mark size=2.5pt}]
        coordinates {(0,0) += (0,16.1) -= (0,0)};
        \addplot[ybar, bar shift=0pt, fill=abEnabled, draw=none,
          error bars/.cd, y dir=both, y explicit,
          error bar style={black, semithick},
        error mark options={rotate=90, black, mark size=2.5pt}]
        coordinates {(1,100) += (0,0) -= (0,16.1)};
        \node[font=\small, anchor=south] at (axis cs:0,21) {$0/20$};
        \node[font=\small, anchor=south] at (axis cs:1,104) {$20/20$};
      \end{axis}
    \end{tikzpicture}\par
    {\footnotesize Separate generations\par
      \smallskip
      Plume: $0/10\rightarrow 10/10$\par
      Two wells: $0/10\rightarrow 10/10$\par
      \smallskip
    Grounding and normalization\\enabled.\par}
  \end{minipage}\hfill
  \begin{minipage}[t]{0.32\textwidth}
    \centering
    {\small\bfseries (c) Normalization}\par
    {\small Accepted yield (\%)}\par
    \smallskip
    \begin{tikzpicture}
      \begin{axis}[abContrast]
        \addplot[ybar, bar shift=0pt, fill=gray!55, draw=none,
          error bars/.cd, y dir=both, y explicit,
          error bar style={black, semithick},
        error mark options={rotate=90, black, mark size=2.5pt}]
        coordinates {(0,45) += (0,20.8) -= (0,19.2)};
        \addplot[ybar, bar shift=0pt, fill=abEnabled, draw=none,
          error bars/.cd, y dir=both, y explicit,
          error bar style={black, semithick},
        error mark options={rotate=90, black, mark size=2.5pt}]
        coordinates {(1,100) += (0,0) -= (0,16.1)};
        \node[font=\small, anchor=south] at (axis cs:0,70) {$9/20$};
        \node[font=\small, anchor=south] at (axis cs:1,104) {$20/20$};
      \end{axis}
    \end{tikzpicture}\par
    {\footnotesize Identical saved decks\par
      \smallskip
      Plume: $9/10\rightarrow 10/10$\par
      Two wells: $0/10\rightarrow 10/10$\par
      \smallskip
    Grounding and synthesis\\enabled.\par}
  \end{minipage}
  \caption{{Effects of grounding, synthesis, and normalization. Each `without' or `with' setting includes
    20 runs (10 per flow benchmark); {error bars are two-sided 95\% Wilson intervals.}
    Panels (a) and (b) report the generated-deck flow ablation. Panel (c) reports a separate
    comparison that reruns the same 20 decks saved from that ablation. Panel (a) measures schema validity; panels (b)
    and (c) measure final acceptance. The enabled components are shown below each panel,
    and runtime repair is disabled.}
  }
  \label{fig:ablation}
\end{figure}
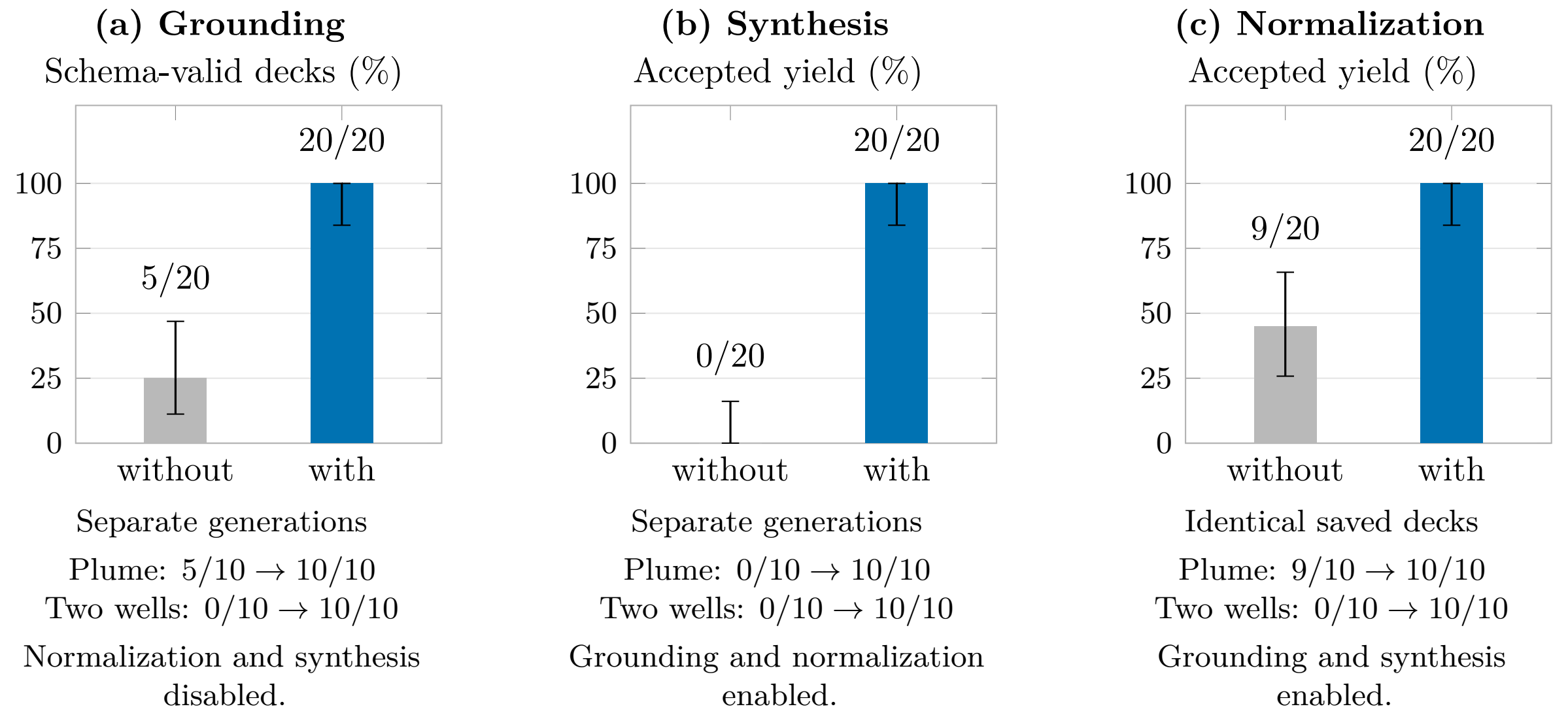

\FloatBarrier
\subsection{Contact capability demonstration}
\label{sec:res:contact}
\label{sec:res:incremental}

{The prompt-to-result T4 run generates a new mesh with $22{,}280$ tetrahedra and $922$
  contact faces and a new input deck, then executes the requested schedule in {$10.9$
  minutes without human intervention}, including deterministic preparation and postprocessing. The simulation
  completes without runtime repair.}

{All 922 faces are classified as sticking, with no opening or sliding, at each of the
  9 saved output times. The contact and hydraulic-aperture checks also pass at each time. The final
  CO$_2$ inventory agrees with the requested $60$ million kg to within $0.378$\,kg. Its change from
  day 30 to day 31 is $5.22\times10^{-8}$\,kg, consistent with injection stopping before monitoring.
  Together, these results satisfy the physics-gate and {input- and output-fidelity requirements}.}

{This demonstration is limited to the observed sticking regime.
The fixed-fault-storage approximation and its supporting sensitivity analysis are discussed
in \Cref{sec:app:contact-storage}.}

\FloatBarrier
\subsection{Paired supplied-deck repair evaluation}
\label{sec:res:repair}
\label{sec:res:selfheal}

{All 20 flow-ablation attempts with complete deterministic preparation finish their
  first simulation execution. Because runtime repair is invoked only after a failure, it is not
  exercised in that ablation. We therefore conduct a separate paired supplied-deck evaluation.}

{In T2, the LLM restores the reference to an existing well-control block or inserts the
  missing relative-permeability table using values stated in the request. Only the named
  attribute or table changes, and both corrected simulations pass the physics gate and all 14
  {input- and output-fidelity checks}. In T3, one attribute edit restores the solid-solver reference.
  The subsequent simulation passes the physics gate and exactly reproduces the baseline in all
  495 comparisons of pressure, stress, and displacement across regions and saved times.}

{All three defective decks stop execution with repair disabled. With repair enabled,
  each defect is corrected by one local edit, after which the simulation is accepted. Both valid
  baseline decks pass in both settings. No repair occurs for the valid baselines, so their input
  decks remain unchanged (\Cref{tab:targeted-repair}).
  \Cref{sec:app:repair-evaluation,sec:app:trajectory-protocol} give the repair protocol and
  trajectory criteria.}

\begin{table}[t]
  \centering
  \caption{Matched supplied-deck repair tests. Each pair starts from identical inputs.
  {\verdictpass{} requires both the physics-gate checks and {all required input- and output-fidelity checks} to pass.}}
  \label{tab:targeted-repair}
  \small
  \begin{tabularx}{\textwidth}{@{}XllX@{}}
    \toprule
    Input & No repair & Repair enabled & Accepted edit or outcome \\
    \midrule
    {Valid two-well flow deck} & \verdictpass{} & \verdictpass{} & No repair edit \\
    Broken well-control reference & Non-pass & \verdictpass{} & One attribute \\
    Missing relative-permeability table & Non-pass & \verdictpass{} & {One table added} \\
    {Valid continuum flow--poromechanics deck} & \verdictpass{} & \verdictpass{} & No repair edit \\
    Broken solid-solver reference & Non-pass & \verdictpass{} & {One attribute} \\
    \bottomrule
  \end{tabularx}
\end{table}

\subsection{Supporting validation tests}
\label{sec:res:guards}
\label{sec:res:correctness}

\subsubsection{\texorpdfstring{{Analytical verification of coupled consolidation}}{Analytical verification of coupled consolidation}}
\label{sec:res:terzaghi}

{A separate, deterministically prepared Terzaghi consolidation test compares
coupled pressure and displacement with an independently evaluated analytical solution
(\Cref{sec:app:terzaghi}). Separate mesh and timestep refinement reduce the largest sampled
errors. With 80 cells and a maximum timestep of $0.0625$\,s, the largest relative errors
at the four sampled times are $0.334\%$ for pressure, $1.070\%$ for displacement,
and $0.820\%$ for loaded-end settlement. TPFA and hybrid mimetic flow give the same
reported errors on this orthogonal mesh. These GEOS runs verify the tested
single-phase coupled configuration; they do not evaluate agentic setup or repair.}

\subsubsection{\texorpdfstring{{Fidelity to the injection schedule}}{Fidelity to the injection schedule}}

{During development, an injection-and-monitoring problem exposed a failure that mass conservation
  alone could not detect. The request specified $2$\,kg/s for $1460$ days followed by monitoring
  without injection until day $3650$. Generated decks instead used $10$ or $5$\,kg/s and continued
  injection. {The simulated increase in stored CO$_2$ matched the mass injected at those rates,
so the mass-balance check passed even though the simulations did not follow the request.}}

{Source synthesis constructs the requested injection period. Input-fidelity checks
  compare the injection rate and injection period specified in the input deck with the
  problem description, while output-fidelity checks assess the saved CO$_2$ inventories
  against the requested injection and monitoring behavior.
  Five end-to-end runs all pass these checks and the physics gate, with a median elapsed time
  of $2.38$ minutes (\Cref{sec:app:shutoff}). These five runs evaluate the implemented response
to the injection-rate and timing errors.}

\subsubsection{\texorpdfstring{{Complementary validation checks}}{Complementary validation checks}}

{A separate seven-defect poromechanics test bypasses pre-solve checks to examine what
  solver execution and the post-solve physics gate detect without input checks
  (\Cref{tab:coverage}).} {Five defective decks abort or
  fail the gate. The negative-solid-density and caprock-permeability cases complete and pass
  because neither changes a checked output enough to violate its criterion at the simulated
  horizon.}

{The solid-density error illustrates why input checks remain necessary: the post-solve
  density check reads fluid density, whereas the complete workflow rejects negative solid
  density before execution. In the caprock case, the sealing layer is assigned the storage-zone
  permeability. This value is admissible in isolation but does not represent
  the requested sealing-layer properties. Detecting the error therefore requires comparing
  each region's assigned permeability with the requested reservoir--caprock configuration.}

{A separate unit-level test supplies scalar and summary quantities directly to the
  physics-gate functions. The functions detect all 15 introduced violations, while the unchanged
  baseline triggers no false alarm among its 14 applicable checks (\Cref{tab:gate}).
  This tests the check functions, not extraction from simulation files.}

\subsubsection{\texorpdfstring{{Reference comparisons}}{Reference comparisons}}

{A comparison with the SPE11B CO$_2$-storage benchmark
\cite{spe11,spe11results} assesses CO$_2$ inventory histories extracted from a rerun
of an expert-authored GEOS deck against those in the official \texttt{geos1} submission.
It uses saved outputs outside LangGraph and does not evaluate agentic input generation or repair.}

{Agreement with the published reference is not uniform
across quantities.} {The extracted mobile CO$_2$ inventory in box~A has a relative trajectory $L_2$
error of $1.6\%$ (\Cref{fig:spe11b}). The other six relative trajectory
errors range from 0.504 to 1.677, and the mean over all seven trajectories is 0.887
(\Cref{tab:e3a}).}

{The continuum flow--poromechanics test compares a modified numerical setup
with its baseline. It increases the initial time step from $10^3$ to $10^6$\,s and also disables
nonlinear line search, which controls damping of Newton updates. Both repair-disabled and
repair-enabled executions complete and pass the physics gate, so neither invokes workflow
repair. In both runs, the maximum normalized field-increment difference from the baseline is $8.57\%$,
exceeding the specified $1\%$ reproduction criterion. They therefore fail the output fidelity
comparison required to reproduce that baseline (\Cref{sec:app:trajectory-protocol}).}

\subsubsection{\texorpdfstring{{Pre-execution escalation}}{Pre-execution escalation}}

{In the tested GEOS build, combining runtime fault generation with an internal well aborts
  before the first time step. Removing runtime fault generation eliminates the reported
  well-subregion error. The alternatives in this setup are to split the fault in the mesh
  beforehand or represent injection by a volumetric source.}

{The separate graph-routing test requires both runtime fault generation and the internal
  well to be retained. The workflow detects the conflict before deck generation and sends the request
  for human review. This tests recognition and escalation of a known incompatibility, rather
  than exhaustion of runtime retries (\Cref{sec:app:compatibility}).}

\subsection{\texorpdfstring{{Portability of input preparation to another multi-physics simulator}}{Portability of input preparation to another multi-physics simulator}}
\label{sec:res:portability}

{In a separate analysis using OpenFOAM, we examine whether predefined construction rules
  help produce complete, runnable simulation inputs for a second simulator. We focus on input
  preparation and do not use {the agentic harness of the GEOS-based framework}. OpenFOAM requires
  a directory of input files rather than a single XML deck.}

{For laminar and turbulent single-mesh dam-break cases, retrieved reference templates
  yield 10/10 cases passing structural and phase-fraction-bounds checks. A separate comparison
  uses prose-only grounding without those templates: adding deterministic synthesis raises the
  laminar pass rate from $10\%$ to $50\%$ and the turbulent pass rate from $10\%$ to $60\%$.}

{We also considered a five-region conjugate-heat-transfer case requiring 56 consistent input files.
  Reference-case grounding supplies the LLM with the complete working case for the same region
  layout. Without deterministic synthesis, LLM generation yields no runnable case either without
  grounding or with reference-case grounding ($0/5$ each); runs with reference-case grounding
  produce approximately 43 files. Here deterministic
  synthesis supplies the missing input files before execution, using templates parameterized by the supplied mesh
  to construct regional dictionaries, fields, and numerical settings. In the GEOS ablation, it
  instead corrects malformed input components and constructs the requested initial physical state.
  With reference-case grounding and synthesis, all five cases run without runtime repair and pass structural and
  heat-transfer checks: bounded temperature, energy-equation convergence, regional continuity,
  and a nonzero thermal gradient.}

\section{Conclusions}
\label{sec:conclusions}

{We set out to determine how task allocation between LLM reasoning and deterministic
  operations can improve reliability in LLM-assisted {multi-physics} simulation. The four conclusions
  below show how our proposed task allocation addresses the failure modes identified in our
  review of prior work.}

\begin{enumerate}[leftmargin=*,itemsep=0.5em]
  \item {\textbf{Grounding and deterministic correction play complementary roles
    in addressing errors in generated input structure.} Grounding guides the LLM toward valid input
    structure through documentation, reference decks, and construction instructions, raising
    generation-stage schema validity from $5/20$ to $20/20$. Deterministic normalization then
    corrects structural omissions and inconsistencies that remain in the generated deck.
    The two therefore act at different stages: grounding informs generation, whereas normalization
    directly modifies the resulting input. In the saved-deck comparison,
    enabling normalization alongside synthesis raises acceptance from $9/20$ to $20/20$ without
    turning a passing result into a non-passing one.}

  \item {\textbf{Schema-valid input still needs construction and checking of a consistent
    physical setup.} Schema-valid inputs do not by themselves establish consistency among the
    physical models, parameters, and numerical configuration. Deterministic synthesis constructs
    input components using pre-specified rules and model-based calculations, while separate checks
    assess physical consistency and agreement with the intended experiment. In the evaluated flow
    cases, adding synthesis raises accepted yield from $0/20$ to $20/20$.}

  \item {\textbf{Controlled local repair can correct configuration errors while preserving
    working inputs.} Paired runs start from identical input decks with repair enabled or disabled.
    Repair corrects each of three seeded defects with one local edit that passes deterministic
    checks, while leaving valid baseline decks unchanged. Repairing the
    continuum flow--poromechanics deck reproduces the baseline pressure, stress, and displacement
    histories across five regions and 33 saved times. These results show that targeted LLM-based
    repair, constrained by deterministic checks, can correct defects in an input deck {without
    regenerating the entire deck}.}

  \item {\textbf{A reusable physics gate and problem-specific fidelity checks provide
    complementary validation.} Our physics gate applies a common check library across flow,
    poromechanics, and contact problems, selecting the applicable {physics} checks for each case.
    {Input- and output-fidelity checks} assess agreement with the problem description and are defined for
    the evaluation at hand. The supporting tests show why both are needed: an incorrect injection
    schedule can conserve mass, and a simulation can pass the physics gate while failing the
    specified baseline-trajectory comparison.}
\end{enumerate}

{The frictional-contact demonstration applies this task allocation to a more complex
multi-physics setting. A valid setup requires consistent choices across flow, poromechanics,
the requested Coulomb contact law, fault hydraulic behavior, fluid-phase ordering, and
mechanical initialization. Deterministic preparation implements these coupled requirements,
and the completed simulation passes the contact, inventory, and schedule checks. This
extension retains LLM interpretation while adding deterministic procedures for the
consistency requirements introduced by the new physics.}

{Within the taxonomy of Lee and Kim~\cite{lee2026agenticcfd}, which distinguishes
assistive, task-autonomous, workflow-autonomous, and research-autonomous systems, our
results support workflow autonomy for user-specified simulations in the tested configurations.
The agentic harness coordinates multiple simulation stages:
it constructs inputs, executes simulations, checks physical consistency and fidelity to the
problem description, and performs constrained repairs without routine human intervention.
Experts still define the scientific objectives, implement construction rules and acceptance
checks, and assess whether the modeling assumptions and simulated scenarios are appropriate
for the practical problem of interest. The contribution is therefore improved reliability of
unattended workflows, not independent scientific investigation. Retaining this task
allocation when the physics or request changes requires implementing and assessing the
corresponding construction rules and checks; our results do not establish autonomous
transfer to unfamiliar configurations.}

\section{Outlook}
\label{sec:limitations}

{Future work has three priorities: improving traceability,
broadening the evidence for workflow reliability, and establishing
{case-specific numerical accuracy.}}

{
  \begin{enumerate}[leftmargin=*,itemsep=0.5em]
    \item \textbf{Improve traceability of inputs, transformations, and assumptions.} The workflow
      retains the request, generated inputs, repair records, simulator logs, and validation reports,
      but does not link {every input value} or modification to its origin. Values can come from
      the request, LLM inference, retrieved examples, deterministic rules, simulator defaults,
      included files, or companion data. A typed intermediate representation and change record
      could preserve these links through generation, normalization, repair, execution, and
      validation. This would allow an auditor to reconstruct how the {final simulation setup} was
      assembled, identify assumptions introduced at each stage, and direct uncertain choices to
      additional checks or human review.

    \item \textbf{Broaden the evidence for workflow reliability.} The two
      flow-ablation benchmarks informed development of the deterministic routines, and the
      supplied-deck repair cases were chosen because they matched repair paths already implemented
      in the framework. Neither group was therefore kept separate from framework development. A future
      held-out evaluation would instead report outcomes for a predefined set of prompts, physical
      problems, and defects withheld during framework development, i.e., evaluated only after the
      deterministic routines, checks, and repair policy were frozen.

      Matched comparisons across LLM backends and agentic systems would assess sensitivity
      to the reasoning component. Complete workflow implementations for OpenFOAM and other
      multi-physics simulators would assess whether the task allocation remains effective beyond
      {the current GEOS workflow and the separate OpenFOAM input-preparation tests}.

    \item {\textbf{Extend numerical verification and calibrate validation tolerances.}
      The analytical test verifies single-phase consolidation, not the multiphase and fault-contact
      configurations of T3 and T4. {Numerical convergence has not yet been established
      for either coupled demonstration.} Their tetrahedral TPFA grids also require assessment of nonorthogonality
      effects. Mesh and timestep error estimates should be used to set tolerances for differences
      between computed results and reference solutions.}
  \end{enumerate}
}

\section*{Acknowledgments}

{Claude Code (using Opus 4 and Fable 5 models) and Codex (using GPT-5 and GPT-6
models) assisted with development of the deterministic tooling
around the LLM agents (normalization and synthesis routines, input checks, physics checks,
and associated tests), and with manuscript drafting, restructuring, and language revision,
under the direction and review of the lead author. The authors retain full responsibility
for the scientific claims, references, and conclusions, including the physics judgment,
assessment of execution failures, and final validation.}

{Claude Code and Codex also assisted with figure-generation code and revisions
to layouts, labels, and captions. Schematics represent the described workflow and case
specifications; charts and plots use recorded experimental results and simulation/reference data.}

\section*{Funding}

{This material is based upon work supported by the Department of Energy under Award Number(s) DE-FE0032793. The authors also acknowledge support by the SUETRI-B Reservoir Simulation Industrial Affiliates Program at
Stanford University.}

\section*{Disclaimer}

{This report was prepared as an account of work sponsored by an agency of the United States Government. Neither the United States Government nor any agency thereof, nor any of their employees, makes any warranty, express or implied, or assumes any legal liability or responsibility for the accuracy, completeness, or usefulness of any information, apparatus, product, or process disclosed, or represents that its use would not infringe privately owned rights. Reference herein to any specific commercial product, process, or service by trade name, trademark, manufacturer, or otherwise does not necessarily constitute or imply its endorsement, recommendation, or favoring by the United States Government or any agency thereof. The views and opinions of authors expressed herein do not necessarily state or reflect those of the United States Government or any agency thereof.}

\section*{Author contributions}

{S.T. designed and implemented the agentic framework, conceived the evaluation studies, 
conducted the numerical experiments, analyzed and interpreted the results, prepared the figures, 
and drafted the manuscript. E.D. provided overall research guidance and supervision, 
and reviewed and revised the manuscript. H.A.T. secured funding and provided technical expertise on the use of the GEOS simulator. All authors approved the final manuscript.}

\section*{Competing interests}

{The authors declare no competing interests.}

\section*{Data availability}

{The SPE11 benchmark specification and the \texttt{geos1} community reference
series are publicly available from the SPE11 project~\cite{spe11,spe11results}.} 




\bibliographystyle{plain}     
\bibliography{references}

\appendix
\crefalias{section}{appendix}
\crefalias{subsection}{appendix}
\crefalias{subsubsection}{appendix}

\section{\texorpdfstring{{Evaluation cases and contact-specific methods}}{Evaluation cases and contact-specific methods}}
\label{sec:cases}

{\Cref{tab:cases} gives the physical specifications of T1--T4 and the SPE11B reference
comparison. The evaluation designs that use these scenarios are mapped in
\Cref{tab:incremental}. For T4, the procedures for establishing initial pressure and stress,
calculating fault permeability, and specifying fracture storage are described in
\Cref{sec:app:contact}.}

\begin{table}[t]
  \centering
  \caption{Physical specifications of the four GEOS scenarios and the SPE11B reference comparison.
    The scenarios are reused across evaluation designs as mapped in \Cref{tab:incremental}.
    SPE11B is an external reference comparison (\Cref{sec:res:correctness}). ``Drive'' is the mechanism that mobilizes CO$_2$; horizons are the
    simulated physical times. All domains use $\lesssim$100{,}000 cells (T4 uses $22{,}280$
  tetrahedra).}
  \label{tab:cases}
  \small
  \begin{tabularx}{\textwidth}{@{}l l >{\raggedright\arraybackslash\hsize=1.35\hsize}X >{\raggedright\arraybackslash\hsize=0.65\hsize}X l@{}}
    \toprule
    \textbf{Case} & \textbf{Domain (km)} & \textbf{Added physics \& structures} & \textbf{Drive} & \textbf{Horizon} \\
    \midrule
    T1 flow-only     & $10\times10\times5$ & compositional CO$_2$--brine, isothermal 50\,$^\circ$C; no wells/faults & emplaced free-gas plume ($S_g\!\approx\!0.6$) & 365\,d \\
    \addlinespace
    T2 two-well flow & $10\times10\times5$ & CO$_2$--brine, isothermal 50\,$^\circ$C; initially 100\% brine; injector \& producer under bottom-hole-pressure (BHP) control & well-pair sweep & 365\,d \\
    \addlinespace
    T3 +poromech.\ \& faults & $10\times10\times5$ & poromechanics; two continuum fault zones; regional-constant materials & \code{SourceFlux}: $500$ million kg over 730\,d, then 182.5\,d monitoring & 912.5\,d \\
    \addlinespace
    T4 +contact      & $8.4\times5\times1.2$ & {frictional Lagrange contact on three faults ($\mu\!=\!0.6$, cohesion 2\,MPa), fixed fracture storage} & two \code{SourceFlux} regions, 30\,d injection + 15\,d monitoring & 45\,d \\
    \addlinespace
    SPE11B           & $8.4\times0.01\times1.2$ & thermal compositional flow; layered facies; no wells/faults & two \code{SourceFlux} sources, each 0.035\,kg/s: years 0--50 and 25--50 & 1000\,yr \\
    \bottomrule
  \end{tabularx}
\end{table}

\subsection{\texorpdfstring{{Contact construction and initialization}}{Contact construction and initialization}}
\label{sec:app:contact}

{The contact construction procedure maps the requested physical laws and parameters
to a shared CO$_2$--brine fluid model and linked fault-storage, hydraulic-aperture,
permeability, and friction models. It selects Phillips brine properties, Span--Wagner
CO$_2$ density, Fenghour CO$_2$ viscosity, and the \code{CO2Solubility} flash model.
The numerical settings below are supplied by this procedure. The framework snapshot is
frozen before the reported prompt-to-result run.}

{The initial state is brine-saturated. Deterministic preparation constructs the
hydrostatic pressure profile by sampling the selected GEOS fluid model in independently constructed
vertical columns with 1\,m and 0.5\,m cell spacing. Both profiles are anchored at $101{,}325$\,Pa
at the $1200$\,m domain top and compared.}

{If $W(z)$ is the integrated saturated weight above elevation $z$, the prescribed
compressive total-stress magnitudes are $101{,}325+W$ vertically, $101{,}325+0.9W$ along
$Y$, and $101{,}325+0.7W$ along $X$, in Pa. Deterministic preparation writes these stresses
and the matching boundary tractions as negative values, following the GEOS sign convention.
Regional Biot coefficients convert the total stresses to effective stresses.}

{For a deck containing a contact solver and surface-element region, normalization constructs
the supporting fault-generation objects from the mesh-exported fault planes. These include
one \code{Rectangle} per fault, the \code{ruptureState} and \code{isFaceSeparable} fields,
a \code{SurfaceGenerator}, and a fracture-generation event. The conformal interfaces split
at $t=0$ before the first solve.}

{Before injection, the workflow performs two mechanical initialization solves and advances the
coupled flow--mechanics model with injection disabled. The resulting fields must satisfy the
initialization, conservation, and contact-law checks.}

{The subsequent 45-day simulation uses
one MPI rank with a direct linear solver, nonlinear tolerance $10^{-8}$, and a maximum one-day
time step. A 100\,kg flow-residual normalization floor permits an absolute mass residual of
$10^{-6}$\,kg in smaller-inventory cells; inventory and phase-volume checks are applied separately.}

\subsubsection{\texorpdfstring{{Hydraulic aperture and fixed storage}}{Hydraulic aperture and fixed storage}}
\label{sec:app:contact-storage}

{Hydraulic aperture follows the Barton--Bandis relation
$a_h=a_0/(1+9\sigma_n/S_0)$, where $\sigma_n>0$ is the magnitude of the compressive normal contact traction,
$a_0\simeq1.5390\times10^{-7}$\,m, and $S_0\simeq46.3837$\,MPa. Permeability is
$a_h^2/12$. Geometric storage aperture remains $a_0$, with unit surface porosity.}

{A five-case storage-sensitivity study started from the equilibrated,
pre-injection state generated by the reported T4 workflow. The runs retained its
mesh, initial fluid and stress states, contact parameters, and
hydraulic-aperture and permeability relations. Each run covered 30 days of injection and
15 days of monitoring.}

{Three cases used fixed fault storage at $0.1$, $1$, and $10$ times the baseline
value. A matched fixed-storage control changed the geometric aperture and porosity while
preserving the baseline initial pore volume and component inventories. A fifth case started
with the same initial pore volume and inventories but allowed fault porosity to vary linearly
with pressure. In this variant, fault storage changes through pressure-dependent porosity
rather than being calculated from the simulated fault deformation. Comparisons at nine
common saved times assessed pressure fields, component inventories, and contact states.}

{Across the four comparisons with the fixed-storage baseline, the largest
relative RMS difference in pressure change was $0.000443\%$, normalized by the baseline
pressure-change RMS within each region. No contact-state classifications changed. Initial
fault pore volumes ranged from $0.281$ to $28.130$\,m$^3$, compared with approximately
$7.56\times10^9$\,m$^3$ of bulk pore volume. These results support fixed storage over this
tested range, with all runs remaining in the sticking regime. They do not test fault opening,
sliding, or fault-storage volumes comparable to bulk storage.}

\section{\texorpdfstring{{Supporting validation evidence}}{Supporting validation evidence}}
\label{sec:appendix-evidence}

\subsection{\texorpdfstring{{Analytical consolidation verification}}{Analytical consolidation verification}}
\label{sec:app:terzaghi}

{
We consider one-dimensional Terzaghi consolidation~\cite{geosterzaghi} in a
$10\times1\times1$\,m saturated column without gravity. The loaded end ($x=0$) is
drained; the opposite end ($x=L=10$\,m) is fixed and impermeable, and lateral strain
is constrained. The material has Young's modulus $E=10^4$\,Pa, Poisson ratio
$\nu=0.2$, porosity $0.3$, permeability $k=10^{-4}$\,m$^2$, and fluid viscosity
$\eta=1$\,Pa\,s. The fluid is incompressible, and a grain bulk modulus of
$10^{27}$\,Pa approximates incompressible grains and a unit Biot coefficient.
From an unloaded state with zero pressure and displacement, compressive traction
rises to $w=1$\,Pa over $10^{-10}$\,s; the first solver step is $10^{-5}$\,s.

In the instantaneous-load, incompressible limit, the independent reference is
\begin{align}
  p(x,t) &= \frac{4w}{\pi}\sum_{m=0}^{\infty}\frac{\sin(\lambda_m x)}{2m+1}
    \exp(-c\lambda_m^2t), & \lambda_m &= \frac{(2m+1)\pi}{2L}, \\
  u(x,t) &= \frac{1}{H}\left[w(L-x)-\int_x^L p(s,t)\,\mathrm{d}s\right],
  & H &= \frac{E(1-\nu)}{(1+\nu)(1-2\nu)},\qquad c=\frac{kH}{\eta}.
\end{align}
The series uses 1,000 terms. We use eight-node hexahedral finite elements for mechanics,
coupled to either TPFA or hybrid mimetic flow. The meshes are uniform, with
$40\times1\times1$ or $80\times1\times1$ cells. For the hybrid mimetic method,
we use the Beir\~ao da Veiga--Lipnikov--Manzini
inner product.
Pressure and axial displacement are compared at saved cell centers and nodes,
respectively, at $t=1,10,30,90$\,s using relative discrete $L_2$ errors;
settlement is the mean axial displacement at the loaded end.

\begin{table}[!htbp]
  \centering

  \caption{Analytical consolidation errors, maximized over the four sampled times,
    not over the complete trajectory. Both flow methods give the values shown
    to the reported precision.}
  \label{tab:terzaghi}
  \small\renewcommand{\arraystretch}{1.15}
  \begin{tabular}{@{}ccrrr@{}}
    \toprule
    Cells & Maximum timestep (s) & Pressure (\%) & Displacement (\%) & Settlement (\%) \\
    \midrule
    40 & 0.25 & 1.052 & 3.315 & 2.673 \\
    80 & 0.25 & 0.993 & 3.089 & 2.387 \\
    80 & 0.0625 & 0.334 & 1.070 & 0.820 \\
    \bottomrule
  \end{tabular}
\end{table}

}

\subsection{\texorpdfstring{{Coverage of input and output checks}}{Coverage of input and output checks}}
\label{sec:app:gate-library}

{The check-library test supplies baseline quantities directly to the physics-gate
functions and changes one quantity at a time. It tests whether each function detects that change,
without reading simulation files or evaluating fidelity to a problem description.}

{\Cref{tab:gate} groups the 15 perturbations by changed quantity. They exercise 14
applicable checks because two perturbations target different conditions within the saturation-bounds
check: a saturation above one and phase saturations that do not sum to one. A violation is counted as
detected if the check returns \verdictfail{} or \verdictwarn{}. None of the checks raises an alarm
on the unchanged baseline.}

\begin{table}[t] \centering \caption{{Direct check-library test: introduced violations and detections by category.
The unchanged baseline raises no alarm among its 14 applicable checks.}}
  \label{tab:gate} \small
  \begin{tabularx}{\textwidth}{@{}X c c@{}} \toprule \textbf{Check
    category} & \textbf{Seeded errors} & \textbf{Caught} \\ \midrule Admissibility (saturation,
    pressure, composition, porosity, density) & 6 & 6 \\ Conservation (CO$_2$ mass balance,
    monotonicity) & 2 & 2 \\ Constitutive (Biot bounds, PVT temperature range, fault slip) & 3 & 3
    \\ Coupling (poroelastic uplift sign, injectivity) & 2 & 2 \\ Numerical (solver convergence,
    stress finiteness) & 2 & 2 \\ \midrule \textbf{Total} & \textbf{15} & \textbf{15} \\
    \bottomrule
  \end{tabularx}
\end{table}

{A separate test introduces seven input defects, one at a time, into a valid continuum
poromechanics deck and runs the solver with pre-solve checks bypassed. This exposes what
solver execution and the post-solve physics gate detect on their own (\Cref{tab:coverage}).
Negative solid density requires an input check because the saved density checked by the gate
is fluid density. {Assigning the storage-zone permeability to the caprock is not
detected by a permeability admissibility check; it requires comparing the region's assigned
permeability with the requested sealing-layer properties.}}

\begin{table}[t] \centering \caption{{Seven input defects tested by direct execution and the post-solve physics gate,
with pre-solve checks bypassed. A gate \verdictpass{} here does not imply workflow acceptance.}}
  \label{tab:coverage} \small
  \begin{tabularx}{\textwidth}{@{}>{\raggedright\arraybackslash}X
    >{\raggedright\arraybackslash}X@{}} \toprule \textbf{Seeded defect} &
    \textbf{Direct solve $+$ post-solve physics gate} \\ \midrule Negative Young's modulus & aborts
    at initialization \\ Poisson ratio $0.95$ (inadmissible) & aborts at initialization
    \\ {Solver references an undefined region} & aborts at initialization \\ Grain modulus
    below drained modulus & aborts at initialization \\ Reference porosity $1.8$
    & {execution completes}; gate \verdictfail{} \\ Caprock at storage-zone permeability (no longer a seal) &
    {execution completes}; gate \verdictpass{}---output matches the baseline run \\ Negative solid density
    & {execution completes}; gate \verdictpass{}---invisible to the fluid-density check \\ \midrule
    Valid baseline deck & {execution completes}; gate \verdictpass{} \\ \bottomrule
  \end{tabularx}
\end{table}

\subsection{\texorpdfstring{{Injection-schedule fidelity}}{Injection-schedule fidelity}}
\label{sec:app:shutoff}

{The prompt requests injection at $2$\,kg/s for $1460$ days, followed by monitoring without
injection until day $3650$. The development failures motivating the test are described in
\Cref{sec:res:guards}.}

{Deterministic synthesis constructs the timed source. Input fidelity checks verify its rate,
component, units, and active interval. At shutoff, the increase in CO$_2$ inventory from its
initial value must agree with the requested $252{,}288{,}000$\,kg injection within $1\%$. At every
saved monitoring time, the change from the shutoff inventory must also remain below $1\%$ of the
requested injected mass.}

{To represent the requested schedule exactly, the workflow ends a simulation time step when
injection stops and saves the CO$_2$ inventory both before injection and at shutoff. Later saved
inventories cover the monitoring period.}

{Each of the five runs starts from the same problem description and independently generates its
mesh and input deck. The largest relative mismatch at shutoff is $6.37\times10^{-7}$, and the
largest relative change in inventory during monitoring is $3.20\times10^{-11}$. The median
wall-clock time for the full workflow, from the problem description through validation, is
$142.72$\,s.}

\subsection{\texorpdfstring{{Contact--well compatibility and escalation}}{Contact--well compatibility and escalation}}
\label{sec:app:compatibility}

{A focused GEOS test combines runtime fault generation
(\code{SurfaceGenerator}) with an internal well (\code{InternalWell}). The input deck passes
schema validation but stops before its first time step
with a well-subregion error. Removing runtime fault generation removes the error, identifying this
feature combination as incompatible in the tested setup. An internal well can instead be used with
a fault already split into the input mesh, or a \code{SourceFlux} can replace the well.}

{A separate {\langgraph{}} routing test requests runtime fault generation and an internal well while
prohibiting either alternative. The feasibility check sends the request to human review before deck
generation. The first test identifies the incompatibility in GEOS; the second verifies the workflow's
pre-execution response.}

\subsection{\texorpdfstring{{SPE11B reference comparison}}{SPE11B reference comparison}}
\label{sec:app:qoi}

{In separate evaluation scripts, we compare CO$_2$ inventory histories extracted from
saved outputs of an expert-authored GEOS SPE11B deck with those from the official
\texttt{geos1} community submission. This comparison does not evaluate LLM-based deck
generation, LangGraph orchestration, or automatic repair.
SPE11B was used during framework development and is not treated as a held-out evaluation.}

\begin{figure}[!htbp]
  \centering
  \includegraphics[width=\textwidth]{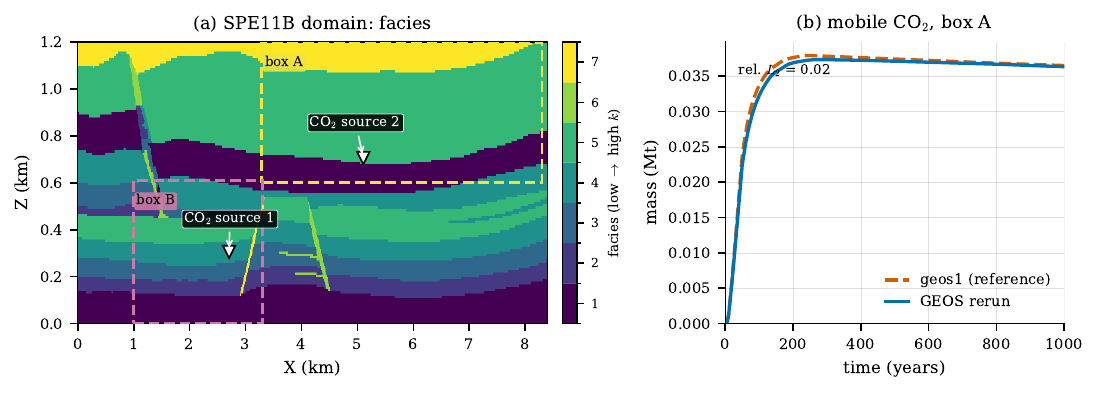}
  \caption{{SPE11B reference comparison. (a) Layered facies, reporting boxes A and B,
and {the two CO$_2$ injection locations in the SPE11B domain}, shown as an
$8.4\,\mathrm{km}\times1.2\,\mathrm{km}$ cross-section.
(b) Mobile CO$_2$ in box~A from the rerun and official \texttt{geos1} submission.
\Cref{tab:e3a} also reports mobile CO$_2$ in box~B, immobile and dissolved CO$_2$ in both boxes,
and CO$_2$ in the seal within box~A.}
  }
  \label{fig:spe11b}
\end{figure}

{\Cref{fig:spe11b} shows the domain and mobile CO$_2$ trajectory in box~A;
\Cref{tab:e3a} reports the full comparison. The table gives relative trajectory $L_2$ errors
over 0--1000 years and, in parentheses, the rerun value at 1000 years divided by the reference
value at the same time.}

{The rerun trajectories are calculated from saved spatial fields at 5-year intervals,
whereas the reference trajectories were calculated within the simulator and interpolated to the
same comparison times. Both simulations use the $840\times120$ SPE11B spatial grid.}

\begin{table}[t]
  \centering
  \caption{{SPE11B inventory comparison against the official \texttt{geos1} submission.
Entries are relative trajectory $L_2$ errors; parentheses give the rerun value at 1000 years
divided by the reference value at the same time. The mean covers the seven reported quantities.}}
  \label{tab:e3a}
  \small
  \begin{tabular}{@{}lcc@{}}
    \toprule
    \textbf{Quantity} & \textbf{Box A} & \textbf{Box B} \\
    \midrule
    Mobile CO$_2$    & 0.016 (1.00) & 0.901 (1.34) \\
    Immobile CO$_2$  & 0.504 (1.10) & 0.700 (1.40) \\
    Dissolved CO$_2$ & 1.113 (2.18) & 1.298 (2.39) \\
    CO$_2$ in seal   & 1.677 (3.24) & --- \\
    \midrule
    \textbf{Mean rel.\ $L_2$ (7 QoIs)} & \multicolumn{2}{c}{\textbf{0.887}} \\
    \bottomrule
  \end{tabular}
\end{table}

\subsection{\texorpdfstring{Held-out evaluation of the optional {critic}}{Held-out evaluation of the optional critic}}
\label{sec:app:critic}

{We evaluated the optional critic separately on 12 matched pairs of valid and deliberately
defective inputs, excluding cases used during development. Each defective input contradicted one
explicit requirement that the existing deterministic checks did not assess. The defective input
and its paired valid input therefore both passed those checks before critic review. These reviews
did not require a GEOS run. Before evaluation, we specified the
performance required to include the critic in the primary claims and accepted-yield comparison:
{detection of at least 10 of 12 contradictions and approval of at least 11 of 12 valid inputs
without edits or human review},
with no edits to clean inputs or to fields outside the identified defect.}

{The critic identified all 12 contradictions. Only 1 of 12 valid inputs was approved unchanged;
the other 11 were preserved and routed to human review (\Cref{tab:critic}). Seven of 12 proposed
corrections to defective inputs passed the existing input checks. One accepted correction also
changed an unrelated statistics-output interval. The critic therefore did not meet the predefined
thresholds and was excluded from the primary claims and accepted-yield comparison.}

\begin{table}[t]
  \centering
  \caption{{Held-out evaluation of the optional critic on 12 matched pairs of valid and
  deliberately defective inputs. Sensitivity is target-specific defect detection; baseline
  specificity is the proportion of valid originals approved unchanged. Intervals are two-sided
  95\% Wilson intervals.}}
  \label{tab:critic}
  \small
  \begin{tabularx}{\textwidth}{@{}Xccc@{}}
    \toprule
    \textbf{Outcome} & \textbf{Count} & \textbf{Rate} & \textbf{95\% Wilson CI} \\
    \midrule
    Target contradiction identified & $12/12$ & $100\%$ & $75.8$--$100\%$ \\
    Valid original approved unchanged & $1/12$ & $8.3\%$ & $1.5$--$35.4\%$ \\
    Defect correction accepted after checks & $7/12$ & $58.3\%$ & --- \\
    Valid original routed to human review & $11/12$ & $91.7\%$ & --- \\
    \bottomrule
  \end{tabularx}
\end{table}

\section{\texorpdfstring{{Framework implementation and evaluation protocols}}{Framework implementation and evaluation protocols}}
\label{sec:app:impl}

\subsection{\texorpdfstring{{Shared state and node interfaces}}{Shared state and node interfaces}}
\label{sec:app:nodes}

{The typed state holds the request, candidate inputs, per-stage status, and inter-node data.
Repair records include retry counters, a simulator-independent error category, the structural
baseline of the first validated deck, previously attempted fixes, and physics feedback.
These records support resumption and prevent repeated unsuccessful edits.}

{\langgraph{} passes the shared state between nodes and executes the transitions
selected by deterministic routing functions. These functions use stage status, error categories,
retry counts, and human-review flags to determine progression, retries, and escalation.
{Deck-validation and runtime-error loops can return directly to deck preparation;
failed postprocessing checks with diagnostic feedback return through the orchestrator when
automatic repair is enabled and retries remain.} Checkpointing preserves the shared
state for workflow resumption.}
{Preparation and validation operate within the existing nodes shown in
\Cref{fig:graph}; they are not additional graph nodes. The generated-deck path constructs
an initial candidate, while the supplied-deck path starts from an existing input.}

\subsection{\texorpdfstring{{Grounding and preparation rules}}{Grounding and preparation rules}}
\label{sec:app:det}

{Grounding uses a {fixed, locally stored copy} of GEOS's auto-generated data-structure documentation.
Per-element tables list attribute names, types, defaults, and descriptions. The XML schema
is a separate input to structural validation. Fixed prompt instructions add cross-element
constraints encountered during framework development; they guide generation but do not verify that
the generated deck satisfies those constraints.}

{Normalization reconciles the deck with the mesh and supporting input structure, including
unit conversion, material references, and cell-block mappings. Synthesis constructs specified
components from extracted parameters. Initial-state and well-control calculations use the
selected fluid model. Files referenced by the input deck are copied into the simulation working
directory before execution.}

{For a declared fault-free problem, fixed mesh-selection rules choose a structured hexahedral
grid and reject unrequested faults. Contact-specific preparation instead uses conformal fault
planes as described in \Cref{sec:app:contact}. Admissibility and input fidelity checks assess
the assembled setup (\Cref{sec:app:bands}).}

{Synthesis records each supplied default's component, field, value, and reason. The routines
return these records through the problem profile when available, and the deck-generation
node stores them in the shared state. Generated XML does not identify whether a value came from
the request or LLM inference, so the log cannot record all deck assumptions. Logging a default
does not require user approval before applying it.}

\subsection{Model configuration}
\label{sec:app:models}

{Deck generation, repair, and meshing use Claude Sonnet~4.6
  (\code{claude-sonnet-4-6}) through an API gateway. Reference decks are retrieved with a local
  embedding index using \code{all-MiniLM-L6-v2}, without an LLM retrieval step.}
{The orchestrator's progress commentary is generated by the locally hosted
  Llama-3 8B model \code{llama3-groq-tool-use:8b}, served through Ollama, and retained in
  the shared state. Workflow routing uses the deterministic rules described in
  \Cref{sec:arch}.}

{Both API-hosted and locally hosted LLMs use a sampling temperature of $0.1$.} {The held-out critic
  study uses Claude Sonnet~4.6 at the same temperature. The flow ablation holds the model choices
  and sampling temperature fixed across configurations.}

\subsection{\texorpdfstring{{Repair implementation and stopping rules}}{Repair implementation and stopping rules}}
\label{sec:app:repair}

{Simulator diagnostics are classified to select a repair route. Deterministic code applies a
prescribed correction when one is available; otherwise, the LLM proposes a correction using
the existing deck and diagnostics. Checks assess required structure and physical constraints;
the preservation mechanisms differ between the supplied- and generated-deck paths.}

{In the supplied-deck path, a proposal must identify exactly which attribute or XML element it changes. An insertion must
specify one existing parent element, and deletion is allowed only for a named element that is
not referenced elsewhere. Before the revised deck is used, deterministic checks assess its XML
schema, input constraints, physical admissibility, and preservation of required content. An
ambiguous or invalid proposal is rejected, leaving the current deck unchanged.}

{The generated-deck path can instead produce revised XML from the existing deck
and diagnostic feedback. Once a baseline is established, a structural-drift check rejects removal
of recorded solver or initial-condition element types, required top-level sections, or all
well regions. It permits additions and attribute-value changes, so input and
post-solve checks remain necessary to assess the corrected setup and results.}

{The general workflow permits at most 20 correction attempts, reduced to 18 for errors
classified as critical or high severity. It can stop earlier when the same error persists after
several different corrections or when further changes no longer make progress. The paired
supplied-deck evaluation uses a stricter limit of two correction rounds after the initial failed
execution (\Cref{sec:app:repair-evaluation}). Thus, the allowed edit forms limit what may change,
and the retry limits bound how many corrections may be attempted.}

{Repair after failed postprocessing checks has a separate limit of two attempts.
Feedback identifies the failed physics or fidelity requirement and its diagnostic quantities.
A correction may change the erroneous initial condition, boundary control, or constitutive
parameter; the revised deck and simulation must pass the required checks.}

\subsection{\texorpdfstring{{Check applicability, tolerances, and verdicts}}{Check applicability, tolerances, and verdicts}}
\label{sec:app:bands}

{The problem profile selects the checks applicable to each simulation. Pre-solve admissibility
checks assess whether the input satisfies implemented simulator and physical constraints, including
bounds and relationships among temperature, elastic moduli, relative-permeability parameters,
well pressures, and solver regions. \Cref{tab:physics} summarizes the {physics} checks by stage;
contact and schedule checks are detailed in \Cref{sec:app:contact,sec:app:shutoff}.}

{Input fidelity compares the assembled setup with the problem description,
  while output fidelity compares the saved results with the requested outcome or a specified
  reference trajectory. One input-fidelity check compares requested permeability with positive
  values in the deck after converting millidarcies or darcies to square meters. The check passes
  when at least one deck value is within a factor of 12 of the requested permeability. This loose
  threshold detects likely copying or unit-conversion errors; it does not verify the regional
  assignment of every permeability value.}

{{Physics-check} tolerances are fixed constants in the implementation and do not vary with mesh
or time-step resolution. The trajectory comparisons use separate tolerances defined in
\Cref{sec:app:trajectory-protocol}. Future mesh- and time-step-refinement studies are needed to
determine how both sets of tolerances should depend on discretization.}

{An evaluated {physics} check returns \verdictpass{}, \verdictwarn{}, or \verdictfail{}.
An explicitly inapplicable check returns \verdictskip{}. Missing fields or an extraction or
array-processing error leave an applicable check unevaluated; the aggregate report records
\verdictincomplete{} when no stronger non-passing verdict applies. Acceptance requires all
applicable checks to pass and at least one {physics} check to be applicable.}

{A \verdictfail{} report with diagnostic feedback can initiate post-solve repair
when enabled. A \verdictwarn{} or \verdictincomplete{} verdict withholds acceptance without
initiating automatic correction. After the permitted retries, any required check that remains
non-passing withholds acceptance. Input- and output-fidelity failures reported by postprocessing
can enter this loop; checks performed by separate evaluation scripts outside \langgraph{}
do not trigger workflow repair.}

\begin{table}[t]
  \centering
  {
    \caption{{Representative pre-solve admissibility and post-solve physics-gate checks.
Applicability depends on the problem profile. {Input- and output-fidelity checks assess agreement with the request or a specified reference.}}}
    \label{tab:physics}
    \small
    \begin{tabularx}{\textwidth}{@{}l l X@{}}
      \toprule
      \textbf{Stage} & \textbf{Category} & \textbf{Representative checks} \\
      \midrule
      Pre-solve & Admissibility & Permeability and elastic-modulus bounds; relative-permeability endpoint consistency; injector BHP above formation pressure and below the estimated fracture pressure \\
      \midrule
      Post-solve & Conservation & Global CO$_2$ inventory balance against estimated injection and production; CO$_2$ mass monotonicity during injection \\
      & Field bounds & Positive absolute pressure and fluid density; saturation and component-fraction bounds and sums; porosity and Biot bounds; PVT temperature range and thermal consistency \\
      & Containment & Free gas in low-permeability seals, or missing seal contrast in a multi-region case containing free gas \\
      & Coupling response & Injection-zone pressure response considered with mass delivery; sign of process-induced vertical displacement \\
      & Numerical checks & Run completion and convergence diagnostics; finite stress and fault displacement jumps (legacy fallback: fault-region displacement) \\
      \bottomrule
    \end{tabularx}
  }
\end{table}

\subsection{\texorpdfstring{{Generated-deck flow ablation}}{Generated-deck flow ablation}}
\label{sec:app:campaign}

{All planned attempts contribute to the reported totals, including those stopped by an
input check before simulator execution. Runtime repair is disabled, so a failed attempt is not
revised and rerun. Within each benchmark, replicate $r$ uses API seed $1000+r$ in all four
configurations to reduce sampling variation when comparing corresponding attempts across
configurations.}

{Both problem descriptions request a fluid-model-based hydrostatic brine background anchored
at $101{,}325$\,Pa at elevation $5000$\,m. T2 additionally requires injector BHP below $1.2$
times initial formation pressure and producer BHP below formation pressure.}

{{Input- and output-fidelity checks} cover the domain, layers, materials, initial state, 365-day duration,
and output schedule. T1 additionally checks plume mobility, upward migration, and closed
CO$_2$ inventory; T2 checks well controls and inventory accounting. T1 does not require the
plume to reach the caprock within 365 days. T2 does not assess fracture-pressure safety because
the problem description supplies no fracture-pressure threshold.}

{{Without synthesis, the configuration with grounding and normalization fails to reproduce
the requested initial pressure and temperature in all 10 plume attempts.} {In the two-well case, 9 attempts} fail the requested
well-to-formation pressure relationship; the remaining attempt stops at the pre-solve pressure
limit. These counts explain the failures summarized in \Cref{sec:res:ablation}.}

\subsection{\texorpdfstring{{Matched normalization comparison}}{Matched normalization comparison}}
\label{sec:app:normalization-diagnostic}

{Each of 20 saved grounded decks is processed twice, once with normalization disabled and
once with it enabled, giving 40 evaluations. Synthesis and the remaining deterministic preparation
are enabled in both runs. Runtime repair is disabled so that any difference can be attributed to
normalization.}

{Each pair uses the same deck, mesh, acceptance criteria, 365-day duration, and 1800-second
simulation timeout. All preparation and initialization failures are counted. A preliminary
two-deck pilot used to finalize the procedure is not included in the comparison.}

{All 11 non-passing runs stop before the requested simulation takes a time step. Nine well
decks omit the regional-pressure calculation needed to construct their well controls. A tenth
well deck also specifies the wrong number of mesh-region entries. The plume deck defines the
same constitutive component both independently and within a compound solid model, causing
initialization to fail.}

{Normalization adds the missing pressure calculation, corrects the mesh-region entries, and
removes the duplicate constitutive definition. Nine pairs pass in both settings, 11 change from
non-pass to pass, and none change from pass to non-pass. The comparison measures normalization
as a whole; it does not isolate the contribution of each correction.}

\subsection{\texorpdfstring{{Paired supplied-deck repair}}{Paired supplied-deck repair}}
\label{sec:app:repair-evaluation}

{Each pair starts from identical copies of an input deck and its referenced files. Runtime
repair is disabled in one copy and enabled in the other; all other processing and acceptance
requirements are identical. The valid input decks are tested in both settings before the defects
are introduced.}

{After a failure, the repair-enabled run first considers a prescribed deterministic correction.
If none applies, the LLM receives the problem description, defective deck, and simulator diagnostics
and proposes one local edit. It is not shown the correct result or given instructions for reversing
the inserted defect. Before another execution, the deterministic checks in \Cref{sec:app:repair}
must approve the revised deck. Once approved, the revised deck is executed without another LLM
call.}

{The two-well tests break an injector control reference or remove a gas
relative-permeability table whose values are in the problem description. Each defect is corrected
with one proposal followed by a second execution. Acceptance requires passing the physics gate
and all 14 {input- and output-fidelity checks}. A separate comparison confirms that content outside the
approved edit remains unchanged.}

{The continuum case couples isothermal CO$_2$--brine flow to Biot poroelasticity on
$12{,}822$ tetrahedra with two 500\,m-thick fault zones. The materials, mesh, loads, and CO$_2$
injection location and rate remain fixed. The defect replaces one solid-solver reference with a
nonexistent name while retaining the correct solver definition.}

{Before breaking the solid-solver reference in the continuum flow--poromechanics deck,
  the valid deck is run once with repair disabled and once with repair enabled. This tests whether
  enabling repair changes a valid input or its result.}

{After the solid-solver reference is broken, the repair-enabled run permits at most
  two correction rounds, allowing no more than three simulation executions including the initial
  failure. Each execution has a 3600-second limit, and {the results are compared using saved VTK fields}.}

\subsection{\texorpdfstring{{Reference-trajectory comparison for the continuum flow--poromechanics case}}{Reference-trajectory comparison for the continuum flow--poromechanics case}}
\label{sec:app:trajectory-protocol}

{The repaired continuum flow--poromechanics simulation and both repair settings
of the numerical-parameter test are compared with the same unmodified baseline. All runs use
the verified initial state, and pressure, stress, and displacement are saved at 33 specified times,
including injection shutoff and the end of monitoring.}

{Acceptance requires a physics-gate \verdictpass{}, agreement with the initial state, finite
mechanical fields, and no tensile effective stress detected by the stress screen. Inventory error
at shutoff and subsequent inventory drift must each remain below $1\%$ of the requested injection.
The trajectory comparison uses the change in each field from that simulation's own initial state,
thereby comparing the evolution during injection and monitoring rather than any initial offset.}

{
For each pressure, stress, or displacement field $q$, the normalized trajectory difference is
\[
  d=\frac{\|\Delta q-\Delta q_{\mathrm{baseline}}\|}
  {\max\{\|\Delta q_{\mathrm{baseline}}\|,\|q_{\mathrm{floor}}\|\}},
\]
where $\Delta q$ is the change from the run's initial state and
$\Delta q_{\mathrm{baseline}}$ is the corresponding change in the unmodified baseline.
Cell-field norms are volume weighted; nodal displacement uses the Euclidean norm.
Denominator floors are constant fields of $100$\,Pa for pressure and stress and
$10^{-6}$\,m for displacement. The 495 comparisons cover five regions, three fields,
and 33 times.
}

{Because the repair should restore only the broken reference, its trajectory must reproduce
the baseline with $d\leq10^{-6}$. The numerical-parameter test deliberately changes time-stepping
and nonlinear-solver settings and uses the less restrictive criterion $d\leq0.01$. These thresholds
measure agreement with the baseline; they do not measure spatial or temporal discretization error.}

{The numerical-parameter test increases \code{initialDt} from $10^3$ to $10^6$\,s and
disables nonlinear line search, retaining 30 Newton iterations, two permitted time-step cuts,
and nonlinear tolerance $10^{-8}$. The saved trajectory is assessed against the baseline
even when execution and physics checks pass. The test runs with runtime repair disabled and
enabled; neither run invokes repair because both complete and pass the workflow checks.
The separate output-fidelity comparison detects their disagreement with the baseline without
triggering workflow repair. Its outcome is reported in \Cref{sec:res:correctness}.}

\end{document}